\documentclass[journal]{IEEEtran}
\usepackage{enumerate}
\usepackage{graphicx}
\usepackage{epsf}
\usepackage{subfigure}
\usepackage{amsmath}
\usepackage{amssymb}
\usepackage{array}
\usepackage{setspace}
\usepackage[amsmath,thmmarks,amsthm]{ntheorem}
\usepackage[ruled,lined]{algorithm2e}
\usepackage{algorithmic}
\usepackage{color}
\usepackage{amsfonts}
\usepackage{dsfont}
\usepackage{xfrac}
\usepackage{breqn}
\usepackage{bbm}
\usepackage{cite}
\usepackage{multirow}
\usepackage{array}
\usepackage{diagbox}

\graphicspath{{Fig/}}
\theoremseparator{.}

\DeclareMathOperator{\tr}{Tr}
\DeclareMathOperator{\diag}{diag}
\DeclareMathOperator*{\argmax}{\arg\!\max}

\graphicspath{{fig/}}

\begin{document}

% paper title
\title{\LARGE Joint Active and Passive Beamforming for Energy-Efficient STARS with Quantization and Element Selection in ISAC Systems}

\author{
Li-Hsiang Shen,~\IEEEmembership{Member,~IEEE},
Yi-Hsuan Chiu
}

\maketitle

\begin{abstract}
	This paper investigates a simultaneously transmitting and reflecting reconfigurable intelligent surface (STARS)-aided integrated sensing and communication (ISAC) systems in support of full-space energy-efficient data transmissions and target sensing. We formulate an energy efficiency (EE) maximization problem that jointly optimizes a dual-functional radar-communication (DFRC)-empowered base station (BS), considering its ISAC-based active beamforming, along with the passive STARS beamforming configurations of amplitudes, phase shifts, quantization levels, and element selection. Furthermore, relaxed/independent/coupled STARS are considered to examine architectural flexibility. To tackle the non-convex and mixed-integer problem, we propose a joint active-passive beamforming, quantization and element selection (AQUES) scheme based on the alternating optimization: Lagrangian dual and Dinkelbach's transformation tackle fractional equations, whereas successive convex approximation (SCA) convexifies the non-solvable problem; Penalty dual decomposition (PDD) framework and penalty-based convex-concave programming (PCCP) procedure solve amplitude and phase-shifts with the equality constraint; Heuristic search iteratively decides the optimal quantization level; Integer relaxation deals with the discrete element selection variables. Simulation results demonstrate that STARS-ISAC with the proposed AQUES scheme significantly enhances EE while meeting communication rates and sensing quality requirements. The coupled STARS further highlights its superior EE performance over independent and relaxed STARS thanks to its reduced hardware complexity. Moreover, AQUES outperforms existing configurations and benchmark methods in the open literature across various network parameters and deployment scenarios.
\end{abstract}


{\let\thefootnote\relax\footnotetext
{Li-Hsiang Shen and Yi-Hsuan Chiu are with the Department of Communication Engineering, National Central University, Taoyuan 320317, Taiwan. (email: shen@ncu.edu.tw and claire90428@gmail.com)}}

\section{Introduction}

\subsection{Backgrounds and Literature Reviews}

	In recent years, the explosive demands and diverse applications in sixth-generation (6G) wireless systems has accelerated the development of integrated sensing and communication (ISAC) technologies \cite{acm, newc1}. With advances in radar and communication systems, the distinct antenna architectures, hardware designs, and operating spectra have begun to converge, emphasizing significant interest in ISAC from both academic and industrial communities. By leveraging shared radio resources and unified hardware platforms of dual-functional radar-communication (DFRC) \cite{isacris22}, ISAC aims for striking a harmonious balance between telecommunications and sensing. Such integration enhances spectral and energy efficiency (EE), reduces hardware redundancy, and alleviates overall power consumption. ISAC is leveraged in diverse applications including autonomous driving, smart cities, robotics, and intelligent homes by enabling simultaneous communication and perception using the same radio signals \cite{isac1, isac3}. Despite these advantages, ISAC faces several practical challenges, such as complex environmental channels, limited sensing range, and coverage blind spots in either dense, dynamic or multi-user scenarios, which should be substantially addressed \cite{newc2}. Moreover, security of ISAC is discussed in the near-field scenario \cite{newc3}.
	
	Reconfigurable intelligent surfaces (RISs) offers a promising solution by dynamically altering the wireless propagation environment, effectively alternating non-line-of-sight (NLoS) channels into virtual line-of-sight (LoS) paths and providing additional degrees of freedom \cite{isac5, isac7, lett, m_dl_ris}. RIS consists of numerous low-cost and energy-efficient meta-atoms, with each capable of independently controlling the phase-shifts and amplitude of incident signals. Unlike traditional active RF chains, RIS modifies the wireless environment passively without complex signal processing, offering a scalable and low-power alternative \cite{m_ai_ris}. Prior research has demonstrated that integrating RIS with other wireless technologies can significantly improve signal strength, mitigate interference, and extend sensing capabilities \cite{ris2, ris3, m_dro}. By establishing alternative propagation paths and enhancing signal reflectivity, RIS can attractively boost both communication and sensing performances. Accordingly, a growing numbers of works have investigated RIS-assisted ISAC systems, with emphasis on joint base station (BS) and RIS beamforming optimization \cite{isacris1, isacris3, isacris11}. Typical RIS-assisted ISAC can be categorized based on their functional focus. The authors in \cite{isacris5, isacris8} aim for enhancing communication performance, relying on direct transmission paths for sensing. For example, the work in \cite{isacris5} optimizes both transmit and receive beamforming along with passive RIS phase shifts in a multiuser setting. In \cite{isacris8}, RIS-assisted ISAC framework tailored for low-power Internet-of-things devices is proposed, where joint beamforming and RIS control is optimized to escalate data rates and maintain sensing reliability. The works of \cite{isacris6, isacris7} target on energy-efficient regimes by optimizing the beamforming and RIS configurations to reduce BS power consumption and enhance environmental sensing performance.

    However, a fundamental limitation of conventional RIS lies in its half-space operation, as it can only reflect signals. To circumvent this issue, the concept of simultaneously transmitting and reflecting RIS (STARS), also termed as STAR-RIS, has emerged \cite{newc4, thz, my_dstar, star1, star2}. STARS can manipulate signals from full-space coverage areas by enabling each element to both transmit and reflect impinging signals, making it particularly attractive in ISAC involving distributed users and targets on different sides of the surface. There exist three different operating protocols of STARS \cite{mag_mode}, i.e., energy splitting (ES), mode switching (MS) and time switching (TS) mechanisms. ES splits the energy per STARS element between the transmission and reflected signals. MS performs a reflection-only or transmission-only structure of each STARS element. While, TS switches all elements to either the reflection or transmission mode at a single timeslot. Note that all elements are considered switched on for full utilizations. It is observed that ES is the most beneficial in providing multicast and multiuser services \cite{mag_mode}. Recent studies have explored STARS for standalone communication \cite{star1,star2} and sensing enhancement \cite{star3}. The integration of STARS into ISAC systems remains relatively unexplored. The works in \cite{star4, star5} have investigated STARS-aided ISAC systems, which focus on specific problem formulations under ideal hardware or simplified models. For example, paper \cite{star4} has proposed minimizing the estimation of lower bound of direction-of-arrival under certain user rate constraints, whereas the authors in \cite{star5} adopt a time switching protocol to achieve alternation between the full-space communication and sensing. Furthermore, the work of \cite{star6} introduces deep reinforcement learning for constant-modulus waveform design. Papers \cite{novastar, novastar1} formulate beamforming problems to jointly optimize user power allocation under certain rate constraint. In the STARS-assisted ISAC systems, the system must jointly optimize communication throughput and sensing accuracy, which often leads to conflicting beamforming and resource allocation objectives. The paper \cite{star7} has proposed an active STARS-assisted ISAC system to address multiplicative fading and extend full-space coverage by jointly designing DFRC beamformers and active STARS configurations, subject to per-target sensing signal quality and power constraints. However, the practical realization of STARS-based ISAC systems introduces several technical challenges, including the high dimensionality of transmission and reflection coefficients, hardware-imposed coupling constraints, and scalability with respect to different user rate demands. Moreover, none of above works consider the practical models of STARS from implementation perspectives, such as quantization, coupled circuit designs and excessive energy utilizations in STARS.

\begin{table*}[!t]
\centering
\small
\setstretch{1.1}
\caption{Literature Comparison}
\begin{tabular}{|l|c|c|c|c|c|c|c|c|c|c|c|c|c|}
\hline
Comparison Items	&	\cite{thz}	&	\cite{my_dstar}	&	\cite{star1}	&	\cite{star2}	&	\cite{star3}	&	\cite{star4}	&	\cite{star5}	&	\cite{star6}	&	\cite{novastar}	&	\cite{novastar1}	&	\cite{star7}	&	\cite{co1, co3}	&	Proposed	\\ \hline \hline
STARS Working Mode	&	ES	&	ES	&	ES, TS	&	ES	&	ES	&	ES	&	TS	&	ES	&	ES	&	MS	&	ES	&	ES	&	ES	\\ \hline
Communication Task	&	\checkmark	&	\checkmark	&	\checkmark	&	\checkmark	&		&	\checkmark	&	\checkmark	&	\checkmark	&	\checkmark	&	\checkmark	&	\checkmark	&	\checkmark	&	\checkmark	\\ \hline
Sensing Task	&		&		&		&		&	\checkmark	&	\checkmark	&	\checkmark	&	\checkmark	&	\checkmark	&	\checkmark	&	\checkmark	&		&	\checkmark	\\ \hline
STARS Circuit Power	&	\checkmark	&		&		&		&		&		&		&		&		&		&	\checkmark	&		&	\checkmark	\\ \hline
Coupled Phase-Shift	&	\checkmark	&	\checkmark	&		&		&		&	\checkmark	&		&	\checkmark	&		&		&		&	\checkmark	&	\checkmark	\\ \hline
Quantization Effect	&	\checkmark	&	\checkmark	&		&		&		&		&		&		&		&		&		&		&	\checkmark	\\ \hline
On-Off State Control	&		&		&		&		&		&		&		&		&		&		&		&		&	\checkmark	\\ \hline
Deployment Evaluation	&		&	\checkmark	&		&		&	\checkmark	&		&	\checkmark	&		&	\checkmark	&		&	\checkmark	&		&	\checkmark	\\ \hline
\end{tabular}
\label{comp_tab}
\end{table*}

\subsection{Challenges and Contributions}

However, the practical STARS-ISAC systems introduce several critical challenges that remain unaddressed. First, the high dimensionality of the optimization space comprising STARS amplitude and phase-shift leads to a significant increase in computational complexity in algorithm design. This algorithm complexity is further exacerbated by hardware-induced coupling effect \cite{co1, co3}. Moreover, most existing studies adopt idealized STARS models assuming continuous and independent control over amplitudes and phase-shifts. The STARS implementations are constrained by quantized control of PIN diodes \cite{thz}, coupled circuit structures, and non-negligible circuital power. These factors might significantly degrade the theoretical gains of ideal cases. Additionally, it is essential to turn off some elements with the worse channel quality to preserve more power for other element to utilize. To this context, it becomes compellingly essential to develop STARS-ISAC frameworks accounting for the quantization, coupling and and energy-aware features of STARS. The comparison table is provided in Table \ref{comp_tab}.

Against this background, we propose a STARS-ISAC architecture that supports joint downlink communication and target sensing in a full-space multi-user environment. A mixed-integer non-convex optimization problem is formulated to maximize the overall EE. To solve this challenging problem, we apply several mathematical transformations and design a structured alternating optimization (AO) algorithm \cite{AO} to iteratively optimize the optimization variables. Additionally, we discuss the on-off states and quantization levels of the STARS elements, highlighting the impact of element adjustments on system power consumption. Most recent researches of STARS-ISAC system focus on improve the system sum-rate, discuss the trade-off between communication and sensing while meeting communication quality requirements and sensing accuracy. Thus, motivated by previous studies, we investigate the joint optimization of the STARS-assisted ISAC system. The main contributions of this paper are elaborated as follows.
\begin{itemize}
    \item To support simultaneous communications and sensing functionality in a full-space coverage area, we conceive an architecture of STARS-aided ISAC operating under the ES protocol. The STARS partitions the entire space into a transmission (${\rm T}$) region for multi-user downlink communication and a reflection (${\rm R}$) region for sensing the target. Three STARS configurations are considered: relaxed STARS and independent as well as coupled T\&R STARS. Relaxed STARS is associated with each element having independent amplitude and phase-shift control, whereas only the sum of T\&R amplitudes is limited for independent STARS. The coupled STARS inherits independent STARS but with additional coupled phase-shifts owing to its coupled hardware architecture. Moreover, quantization effect is considered to align with the hardware implementation.

    \item We formulate an EE maximization problem by jointly optimizing the active ISAC beamforming at BS and the passive beamforming at STARS including the configurations of amplitudes and phase-shifts of T\&R parts, the selected on-off states of each element, and the quantization levels. The constraints guarantee the minimum user rate and sensing signal quality requirements as well as power budget at BS and at STARS.

	\item To solve the original mixed-integer and non-convex optimization problem, we propose a joint active-passive beamforming, quantization and element selection (AQUES) scheme based on alternating optimization: Lagrangian dual and Dinkelbach's transformation are utilized for tackling fractional problems, whereas successive convex approximation (SCA) aims for convexifying the problem; Penalty dual decomposition (PDD) framework along with penalty-based convex-concave programming (PCCP) procedure solves STARS amplitude and phase-shifts; Heuristic search decides the discrete quantization level; Integer relaxation deals with the STARS element selection.

	\item The proposed STARS-ISAC architecture is evaluated through simulations under different system settings, including transmit power, different numbers of users/antennas/elements, and quantization levels as well as deployment scenarios. The results validate that the coupled STARS achieves the highest EE performance compared with other STARS architectures, owing to its simplified hardware structure. Moreover, the proposed AQUES scheme for STARS-ISAC also accomplishes the highest EE against the cases with the only optimized amplitude/phase-shifts and time/mode switching mechanisms as well as against the benchmarks using centralized reinforcement learning, conventional beamforming, and heuristic method.
\end{itemize}
    
    {\it Notations:}  We define bold capital letter $\mathbf{A}$ as a matrix and bold lowercase letter $\mathbf{a}$ as a vector. The operations $\mathbf{A}^H$, $\mathbf{A}^T$, and $\mathbf{A}^{-1}$ represent the Hermitian, transpose and inverse of $\mathbf{A}$. $\diag(\cdot)$ indicates a diagonal matrix. $\mathfrak{R}\{\cdot\}$ represents the real part of a complex parameter. $\dot x$ is conjugate operation given complex variable $x$. 
%${\rm vec}(\mathbf{A})$ vectorizes the matrix $\mathbf{A}$. 
$\otimes$ indicates Kronecker product. $\preceq$ is elementwise comparison of "less than" or "equal to". $[\mathbf{A}]_{mn}$ is the $(m,n)$-th element of $\mathbf{A}$, whereas $[\mathbf{a}]_{m}$ indicates the $m$-th element of $\mathbf{a}$. $\left\lceil \cdot \right\rceil$/$\left\lfloor \cdot \right\rfloor$ are ceiling/floor functions. $\mathbb{E}[\cdot]$ indicate the expectation operation. $|\cdot|,\lVert \cdot \rVert, \lVert \cdot \rVert_F$, and $\lVert \cdot \rVert_{\infty}$ denote the absolute value, the 2-norm, Frobenius norm, and infinity norm, respectively.
 
%The rest of this paper is organized as follows. Section \ref{sec_sys} describes the system model and problem formulation. The proposed algorithm is elaborated in Section \ref{sec_alg}. Section \ref{sec_sim} offers simulation results, whereas the conclusions are drawn in Section \ref{sec_con}.

\begin{figure}[t]
  \centering
  \includegraphics[width=3in]{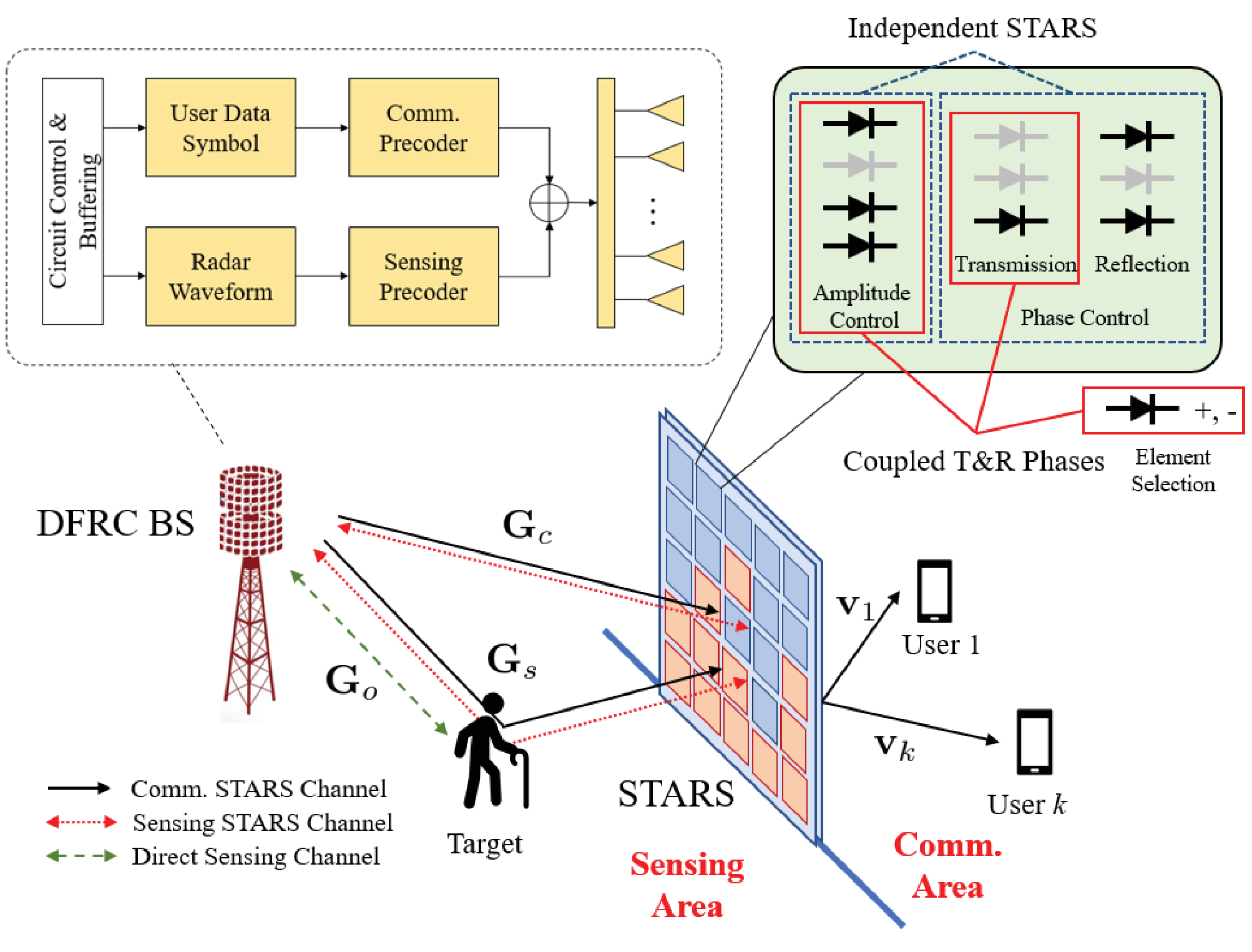}
  \caption{The architecture of DFRC-enabled BS in STARS-ISAC systems.}
  \label{scene}
\end{figure}

%\begin{figure}[t]
%  \centering
%  \includegraphics[width=3.5in]{antenna}
%  \caption{DFRC-enabled hybrid beamforming at BS.}
%  \label{ant}
%\end{figure}
%
%\begin{figure}[t]
%  \centering
%  \includegraphics[width=3.5in]{STARRIS}
%  \caption{Architecture of STARS.}
%  \label{star}
%\end{figure}

\section{System Model and Problem Formulation}
\label{sec_sys}

\subsection{System Model}

\subsubsection{System Architecture and Channel Models}

The STARS-aided ISAC is illustrated in Fig. \ref{scene}, where the BS empowered by DFRC equipped with $N$ transmit antennas simultaneously detects the sensing target and communicates with $K$ downlink users, associated with the respective sets of $\mathcal{N}=\{1,...,N\}$ and $\mathcal{K}=\{1,...,K\}$. Note that the mono-static sensing accomplished by DFRC is considered, i.e., the BS takes care of both transmit sensing signals and detection of received echoes\textsuperscript{\ref{note1}}\footnotetext[1]{Note that DFRC is a functional unit integrating both radar sensing and communications, and is considered one of the ISAC techniques. However, other components can also enable ISAC applications, e.g. bi-static sensing \cite{isac3} employs a single transmit waveform at the BS (without radar functions) and performs detection at the receiving user ends. \label{note1}}. We consider a STARS with $M$ elements with its element set as $\mathcal{M}=\{1,..., M\}$. The direct link from the BS to the users is unavailable owing to potential blockages. The STARS partitions the entire space into two parts, namely the transmission (${\rm T}$) area where the communication users are located and the reflection (${\rm R}$) area where the sensing target is located\textsuperscript{\ref{note2}}\footnotetext[2]{As a future extension, incorporating single-/multi-target sensing and multiusers on both sides of STARS can establish a more general framework. However, these scenarios require complex ISAC channel modeling and problem reformulation with additional constraints under the assumption of presence of the direct links. Accordingly, the detailed analysis is required, and the algorithms must be redesigned to accommodate the generalized frameworks. \label{note2}}. We consider the Rician fading channel model between the BS and the STARS unaffected by the target as $\mathbf{G}_c \in \mathbb{C}^{M \times N}$
\begin{align}
	\mathbf{G}_c \!=\! \sqrt{ h_0 d^{-\alpha_0} }
\left( \sqrt{\frac{\beta_0}{\beta_0 \!+\! 1}} \mathbf{G}_c^{\rm LoS} \!+\! \sqrt{\frac{1}{\beta_0 \!+\! 1}} \mathbf{G}_c^{\rm NLoS} \right),
\label{4}
\end{align}
where $h_0$ is the reference pathloss at 1 meter, $d$ is the distance of BS-STARS, and $\alpha_0$ is the pathloss exponent. Notation of $\beta_0$ is the Rician factor indicating the portion of LoS/NLoS path component $\mathbf{G}_c^{\rm LoS}$/$\mathbf{G}_c^{\rm NLoS}$. The LoS component $ \mathbf{G}_c^{\rm LoS}$ can be defined as the array response vector as
\begingroup
\allowdisplaybreaks
\begin{align}
\mathbf{G}_c^{\rm LoS}
	& = 
\begin{bmatrix}
1,e^{-j\frac{2\pi}{\lambda} d_s \sin\varphi_{s}}, ...,e^{-j\frac{2\pi}{\lambda}(M-1) d_s \sin\varphi_{s}}
\end{bmatrix}^{T} \notag \\
	& \cdot
\begin{bmatrix}
1,e^{-j\frac{2\pi}{\lambda} d_b \sin\varphi_{b}}, ...,e^{-j\frac{2\pi}{\lambda}(N-1) d_b \sin\varphi_{b}}
\end{bmatrix}.
\label{5}
\end{align}
\endgroup
Notation of $\lambda$ indicates the frequency wavelength, and $d_{s/b}$ denotes the inter-antenna and inter-element separation, respectively. $\varphi_{s/b}$ represents the angle-of-arrivals (AoA) of STARS and angle-of-departures (AoD) of BS, respectively. It is assumed that the AoA and AoD at BS/STARS are the same due to channel reciprocity in LoS path. Note that $\mathbf{G}_c^{\rm NLoS}$ follows independent and identically distributed Rayleigh fading. Following \eqref{4}, the channel vector from the STARS to user $k$ is denoted as $\mathbf{v}_{k} \in \mathbb{C}^{N \times 1 }$ where the LoS path is
\begin{align} \label{6}
	\mathbf{v}_k^{\rm LoS} = \left[
1, e^{-j\frac{2\pi}{\lambda} d_s \sin\varphi_{st}}, ..., e^{-j\frac{2\pi}{\lambda}(M-1) d_s \sin\varphi_{st}} \right]^{T}
\end{align}
where $\varphi_{st}$ stands for the AoD of STARS ${\rm T}$-side and their distance of STARS-user $k$ is $d_k$. We define the channels of BS-target and target-STARS as $\mathbf{g}_{s}$ and $\mathbf{r}_{s}$, respectively, where their LoS/NLoS components following \eqref{4}, \eqref{5}, and \eqref{6} are denoted as $\mathbf{g}_{s}^{\rm LoS}\in\mathbb{C}^{N \times 1},\mathbf{r}_{s}^{\rm LoS}\in\mathbb{C}^{M \times 1}, \mathbf{g}_{s}^{\rm NLoS}\in\mathbb{C}^{N \times 1}$, and $\mathbf{r}_{s}^{\rm NLoS}\in\mathbb{C}^{M \times 1}$. The distances of BS-target and target-STARS are $d_{s_1}/d_{s_2}$. Note that $\mathbf{G}_s = \mathbf{g}_{s} \cdot \mathbf{r}_{s}^{H}$ is the cascaded channel of BS-target-STARS. We neglect the similar definitions of the remaining parameters for simplicity. Note that perfect channel information is acquired\textsuperscript{\ref{note3}}\footnotetext[3]{Since STARS can be a passive architecture, imperfect channel information with estimation errors or unavailability might occur. The imperfection will degrade ISAC performances. There exist some potential solutions to address the above mentioned issues: (1) Consider the statistical channel models, the related robust optimization techniques \cite{m_dro} can be applied to maximize and guarantee the worst-case performance within the feasible regions. (2) The learning-based mechanisms are employed to deal with complex and dynamic wireless environments, which require only the performance feedback from users \cite{m_ai_ris}. However, both methods should be redesigned in the proposed STARS-ISAC systems as the future potentials.\label{note3}}.

\subsubsection{STARS Model}
Each STARS element adjusts the amplitude $\boldsymbol{\beta}_{\mathcal{Y}} = [ \beta_{\mathcal{Y},1},..., \beta_{\mathcal{Y},M} ]^T$ and phase-shifts $\boldsymbol{\phi}_{\mathcal{Y}} = [ \phi_{\mathcal{Y},1},..., \phi_{\mathcal{Y},M} ]^T$ with $\mathcal{Y}\in\{\rm T,R\}$ for transmission and reflection, respectively. In particular, the independent STARS follows the energy conservation given by
\begin{align} \label{STAR_amp}
	\beta_{{\rm T},m}^2 + \beta_{{\rm R},m}^2 = 1,  \ \forall m \in \mathcal{M}.
\end{align}
Note that the relaxed STARS follows independent amplitude constraint as $|{\theta}_{\mathcal{Y},m}|={\beta}_{\mathcal{Y},m}$. For independent STARS, each phase-shift can be determined independently as $\boldsymbol{\Theta}_{\rm T} = \diag(\boldsymbol{\theta}_{\rm T}) = \diag([\theta_{{\rm T},1},...,\theta_{{\rm T},M}]^T)$ and $\boldsymbol{\Theta}_{\rm R} = \diag(\boldsymbol{\theta}_{\rm R}) = \diag([\theta_{{\rm R},1},...,\theta_{{\rm R},M}]^T)$ where $\theta_{{\rm T},m} = \alpha_m \beta_{{\rm T}, m} e^{j \phi_{{\rm T},m}},\theta_{{\rm R},m}= \alpha_m \beta_{{\rm R}, m} e^{j \phi_{{\rm R},m}}$ where $\phi_{{\rm T},m}, \phi_{{\rm R},m}\in [0,2\pi)$. The notation $\alpha_m\in\{0,1\}$ indicates the state per STARS element, where $\alpha_m=1$ denotes "on" state and $\alpha_m=0$ indicates "off" state. The corresponding vector of $\alpha_m$ is denoted as $\boldsymbol{\alpha} = [\alpha_1,...,\alpha_M]^T$. As for the low-cost STARS, phase-shifts are coupled \cite{co1}, termed as \textit{coupled T\&R phase-shift} model, where the constraint is given by
\begin{align} \label{STAR_ph}
	\cos ( \phi_{{\rm T},m} - \phi_{{\rm R},m} ) = 0,  \ \forall m \in \mathcal{M}.
\end{align}
That is, $\phi_{{\rm T},m} = \phi_{{\rm R},m} \pm \frac{\pi}{2}$. Inspired by \cite{thz}, the power consumption of STARS with on-off states can be given by
\begin{align} \label{power_star}
	P^{a}_{{\rm ST}} = \sum_{m\in \mathcal{M}} \alpha_m N_{\rm PIN}^{a} P_{\rm PIN} + P_{\rm CIR}.
\end{align}
where $a\in \{re, in,co\}$ indicates the types of relaxed ($re$), independent ($in$) or coupled ($co$) STARS, associated with the average numbers of the required PIN diodes as $N_{\rm PIN}^{re} = \lceil 2 \log_2 L^{re}_{\beta} + 2\log_{2} L^{re}_{\phi} \rceil$, $N_{\rm PIN}^{in} = \lceil \log_2 L^{in}_{\beta} + 2\log_{2} L^{in}_{\phi} \rceil$ and $N_{\rm PIN}^{co} = \lceil \log_2 L^{co}_{\beta} + \log_{2} L^{co}_{\phi} + 1 \rceil$, respectively. The notation $P_{\rm PIN}$ indicates the operating power consumption of each STARS element, whilst $P_{\rm CIR}$ indicates the static circuit power consumption of STARS. The quantization levels of amplitudes and phase-shifts are denoted as $L^a_{\beta}$ and $L^a_{\phi}$, respectively.

%In this work, we consider a single-sided STARS\textsuperscript{\ref{note0}}\footnotetext[1]{The double-sided STARS \cite{?} is capable of supporting impinging signal arrival at both sides of surfaces. However, asymptotic closed-forms are unavailable. Moreover, it requires complex interference control and circuital processing, which can be left as future work. \label{note0}}, i.e., the impinging signals can only arrive at one side of STARS. We consider that sensing object is located at the reflection region of STARS, whilst the communication users locate at transmission region, as shown in Fig. \ref{scene}. 

%Note that the reflected sensing pulses from the object within the transmission region cannot be successfully received at BS. 

\subsubsection{Transmit ISAC Signal}
To carry out ISAC, the BS transmits the beamformed ISAC signals at time $t$ as $\mathbf{x} (t) =  \sum_{k \in \mathcal{K}} \mathbf{w}_{c,k}  c_k(t) + \mathbf{W}_{s} \mathbf{s}(t)$,
%\begin{align}
%	\mathbf{x} (t) =  \sum_{k \in \mathcal{K}} \mathbf{w}_{c,k}  c_k(t) + \mathbf{W}_{s} \mathbf{s}(t),
%\end{align}
where $\mathbf{w}_{c,k}\in \mathbb{C}^{N \times 1}$ and $\mathbf{W}_{s} \in \mathbb{C}^{N \times N}$ stand for beamforming vectors for communications and sensing, respectively. Symbols of $\mathbf{c}(t)=[c_1(t),...,c_K(t)]^T \in \mathbb{C}^{K \times 1}$ and $\mathbf{s}(t) \in \mathbb{C}^{N \times 1} $ are data streams for users and the dedicated sensing signal for detecting the target, respectively. Then the covariance of transmit signals is denoted as $\mathbf{R}_w = \mathbb{E}[ \mathbf{x} (t) \mathbf{x} (t)^H] = \sum_{k\in \mathcal{K}} \mathbf{w}_{c,k} \mathbf{w}_{c,k}^H + \mathbf{W}_s \mathbf{W}_s^H$. We consider $\mathbb{E}[\mathbf{c}(t) \mathbf{c}(t)^H] = \mathbf{I}_K$, $\mathbb{E}[\mathbf{s}(t) \mathbf{s}(t)^H] = \mathbf{I}_N$ and $\mathbb{E}[\mathbf{c}(t) \mathbf{s}(t)^H] = 0$. Accordingly, the power constraint for DFRC-enabled BS is represented by
\begin{align}
	\tr (\mathbf{R}_w ) = \tr\left( \sum_{k\in \mathcal{K}} \mathbf{w}_{c,k} \mathbf{w}_{c,k}^H + \mathbf{W}_s \mathbf{W}_s^H \right) \leq P_{th},
\end{align}
where $P_{th}$ is the available power for precoding at BS. The received communication signal of user $k$ is expressed as
\begin{align}
	y_k^{c}(t) &= \boldsymbol{\theta}_{\rm T}^T \mathbf{H}_{k} \mathbf{w}_{c,k} c_k(t)  + \sum_{k'\in \mathcal{K} \backslash k}  \boldsymbol{\theta}_{\rm T}^T \mathbf{H}_{k} \mathbf{w}_{c,k'} c_{k'}(t) \notag
	 \\
	 & + \boldsymbol{\theta}_{\rm T}^T \mathbf{H}_{k} \mathbf{W}_{s} \mathbf{s}(t) + n_k,
\end{align}
where $n_k$ is the additive white Gaussian noise (AWGN) of user $k$. Notation of $\mathbf{H}_{k} = \diag(\mathbf{v}_k) \mathbf{G}$ indicates the cascaded channel from BS-STARS and from STARS-user, where $\mathbf{G}$ consists of BS-STARS link unaffected by the target and that affected by the target, i.e., $\mathbf{G} = \mathbf{G}_c + \mathbf{G}_s$. Therefore, the received signal-to-interference-plus-noise ratio (SINR) for user $k$ can be expressed as
\begin{align}
	\gamma_{k}^{c} = \frac{ \lVert \boldsymbol{\theta}_{\rm T}^T \mathbf{H}_{k} \mathbf{w}_{c,k} \rVert^2}{ \sum_{k'\in \mathcal{K} \backslash k} \lVert \boldsymbol{\theta}_{\rm T}^T \mathbf{H}_{k} \mathbf{w}_{c,k'} \rVert^2 + \lVert \boldsymbol{\theta}_{\rm T}^T \mathbf{H}_{k} \mathbf{W}_{s} \rVert^2 + \sigma_k^2},
\end{align}
where $\sigma_k^2$ is the noise power of user $k$. The received sensing echo signal from the target is given by
\begin{align}
	\mathbf{y}_s(t) = \mathbf{H}_s \mathbf{x} (t) + \mathbf{n}_s(t),
\end{align}
where $\mathbf{H}_s = \alpha_0 \mathbf{g}_s \mathbf{g}_s^H + \alpha_0^2 \mathbf{G}^H \boldsymbol{\Theta}_{\rm R} \mathbf{G} \in \mathbb{C}^{N \times N}$ is the effective sensing channel, where $\alpha_0$ denotes the absorption coefficient of the target surface. Notation of $\mathbf{n}_s(t)$ indicates noise at BS. After receiving the signal, the receive beamforming $\mathbf{u}_{s} \in \mathbb{C}^{N \times 1}$ is employed for detecting the echo bouncing back to BS. The corresponding sensing SINR can be given by
\begin{align}
	\gamma^{s} = \frac{ \lVert \mathbf{u}_{s}^H \mathbf{H}_{s} \mathbf{W}_{s} \rVert^2}{ \sum_{k\in \mathcal{K}} \lVert \mathbf{u}_{s}^H \mathbf{H}_{s} \mathbf{w}_{c,k} \rVert^2  + \sigma_s^2 \lVert \mathbf{u}_{s}\rVert^2}.
\end{align}
Moreover, sensing is additionally characterized by interference-to-noise ratio (INR) to alleviate the sensing interferences as
\begin{align}
	\gamma^s_{I}  = \frac{\sum_{k\in \mathcal{K}} \lVert \mathbf{u}_{s}^H \mathbf{H}_{s} \mathbf{w}_{c,k} \rVert^2}{\sigma_s^2 \lVert \mathbf{u}_{s}\rVert^2}.
\end{align}
The corresponding ergodic rate for communication part is $R_k = \log_2 \left( 1+ \gamma_{k}^{c} \right)$, whereas the total rate is denoted as $R_t=\sum_{k\in \mathcal{K}} R_k$. Furthermore, the rate-dependent power consumption is modeled as
\begin{align}
	P = \tr (\mathbf{R}_w) + \xi R_t + P_{\rm BS} + P^{a}_{{\rm ST}},
\end{align}
where $\xi$ is a constant incurred by the modulation and coding/decoding schemes and fronthaul/backhaul processing \cite{thz}, and  $P_{\rm BS}$ is BS static power consumption.

\subsection{Problem Formulation}

We aim for maximizing the system EE by optimizing the parameters of ISAC transmit beamforming $\{\mathbf{W}_s,  \mathbf{w}_{c,k}\}$ and receiving beamforming $\mathbf{u}_s$, STARS configurations $\mathbf{\Theta}_{\mathcal{Y}}, \forall \mathcal{Y}\in\{\rm T, R\}$, quantization level $L_{\mathcal{X}}, \forall \mathcal{X}\in\{\beta, \phi \}$, as well as the STARS element selection $\boldsymbol{\alpha}$. The problem is formulated as
 \begingroup
 \allowdisplaybreaks
 \begin{subequations} \label{problem_T}
\begin{align}
    &\mathop{\max}\limits_{\substack{  \mathbf{w}_{c,k}, \mathbf{W}_{s}, \mathbf{u}_s, \boldsymbol{\Theta}_{\mathcal{Y}}, L_{\mathcal{X}}, \boldsymbol{\alpha} }} \quad  
    \frac{R_t}{ P} \label{obj_D2} \\
     \text{s.t. } & \quad \boldsymbol{\Theta}_{\mathcal{Y}} \in \mathcal{F}_\Theta, \ \forall \mathcal{Y} \in \{\rm R, T \}, \label{con1}\\
     &\quad \tr (\mathbf{R}_w) \leq P_{th}, \label{con2}\\
     &\quad P^{a}_{\rm ST} \leq P_{th}^{a}, \ \forall a\in\{re, in, co\}, \label{con3}\\
     &\quad R_k \geq R_{k,th}, \ \forall k\in \mathcal{K}, \label{con4}\\
     &\quad \gamma^s \geq \gamma_{th}^{s}, \label{con5} \\
     &\quad \gamma^s_{I} \leq \gamma_{I,th}^{s}, \label{con6} \\
     &\quad L^{a,min}_{\mathcal{X}} \leq L^{a}_{\mathcal{X}} \leq L^{a,max}_{\mathcal{X}}, \notag \\
     &\qquad\qquad\qquad \forall \mathcal{X}\in \{\beta, \phi\}, \forall a\in\{re, in, co\}, \label{con7}\\
     &\quad \alpha_{m} \in \{0,1 \}, \ \forall m\in\mathcal{M}. \label{con8}
\end{align}
 \end{subequations}
  \endgroup
In problem $\eqref{problem_T}$, the constraint $\eqref{con1}$ indicates the STARS configuration constraints for amplitude and phase-shifts in the feasible domain $\mathcal{F}_\Theta$ depending on STARS types. $\eqref{con2}$ exhibits the DFRC-enabled BS transmit power constrained by $P_{th}$. In $\eqref{con3}$, power consumption of STARS should be lower than $P^{a}_{th}$ for type-$a$ STARS. Constraint $\eqref{con4}$ guarantees the minimum user rate requirement $R_{k,th}$. The sensing SINR should be higher than $\gamma_{th}^s$ in $\eqref{con5}$, whilst INR should be constrained by $\gamma_{I,th}^s$ in $\eqref{con6}$. Owing to hardware limits and performance guarantee, quantization levels of STARS should be within the range of $[L_{\mathcal{X}}^{a,min}, L_{\mathcal{X}}^{a,max}]$ for both amplitude and phase-shifts in \eqref{con7}. The binary on-off state of each element is shown in $\eqref{con8}$. We can observe that it is difficult to directly solve the non-convex and non-linear problem $\eqref{problem_T}$. To circumvent this issue, we employ the AO technique to iteratively obtain the suitable solutions.

\section{Proposed AQUES Scheme} \label{sec_alg}

\subsection{Problem Transformation}
As EE problem performs a fractional programming problem, we firstly transform it with an auxiliary parameter $\eta$ into
 \begingroup
 \allowdisplaybreaks
 \begin{subequations} \label{problem_T2}
\begin{align}
    &\mathop{\max}\limits_{ \mathbf{w}_{c,k}, \mathbf{W}_{s}, \mathbf{u}_s, \boldsymbol{\Theta}_{\mathcal{Y}}, L_{\mathcal{X}}, \boldsymbol{\alpha}, \eta} \quad  
    \eta \\
     \text{s.t. } 
     & \quad  R_t - \eta P \geq 0, \label{con0} \\
     & \quad \sum_{k\in \mathcal{K}} \mathbf{w}_{c,k}^H\mathbf{w}_{c,k} + \lVert \mathbf{W}_{s} \rVert_F^2 \leq P_{th}, \label{con0_1} \\
     & \quad \eqref{con1}, \eqref{con3}-\eqref{con8}.
\end{align}
 \end{subequations}
  \endgroup
The transformed constraint $\eqref{con0_1}$ is equivalent to the inequality in $\eqref{con2}$. For a tractable expression, we further rewrite the constraint $\eqref{con0}$ in an alternative form as
\begin{align} \label{con_r}
	R_t - \eta P \geq 0  \Leftrightarrow 
	(1-\eta \xi) R_t & - \eta P(\mathbf{W}) \geq 0,
\end{align}
where $P(\mathbf{W}) =  \sum_{k\in\mathcal{K}} \mathbf{w}_{c,k}^H \mathbf{w}_{c,k} + \lVert \mathbf{W}_s \rVert_F^2 + P_{\rm BS} + P^{a}_{{\rm ST}} $. Now, the objective function is readily solved by just maximizing $\eta$, whereas the remaining parameters and constraints are tackled by AO below.

\subsection{Solution of Active ISAC Beamforming}
The subproblem for ISAC transmit/receiving beamforming with respect to (w.r.t.) $\{\mathbf{w}_{c,k}, \mathbf{W}_{s}, \mathbf{u}_s\}$ can be formulated as
 \begingroup
 \allowdisplaybreaks
% \begin{subequations} \label{problem_Beam}
\begin{align} \label{problem_Beam}
    & \mathop{\max}\limits_{\substack{   \mathbf{W} \in \{ \mathbf{w}_{c,k}, \mathbf{W}_{s} \}, \\ \mathbf{u}_s, \eta}  }  
    \eta \quad \text{  s.t. }
     \eqref{con4}, \eqref{con5}, \eqref{con6}, \eqref{con0_1}, \eqref{con_r}.
\end{align}
% \end{subequations}
  \endgroup
We exploit the Lagrangian dual transform \cite{my_dstar} to $R_t$ in $\eqref{con_r}$ as
\begin{align} \label{r_Lan}
	R_t = \sum_{k\in\mathcal{K}}  \left[ \log_2(1+ \bar{\gamma}_k^c) - \bar{\gamma}_k^c + \frac{ \left( 1+\bar{\gamma}_k^c \right) A_k(\mathbf{W})}{A_k(\mathbf{W}) + B_k(\mathbf{W})} \right],
\end{align}
where $\bar{\gamma}_k^c$ is the SINR obtained at previous iteration, and pertinent parameters are defined as $A_k(\mathbf{W}) =  \lVert \boldsymbol{\theta}_{\rm T}^T \mathbf{H}_{k} \mathbf{w}_{c,k} \rVert^2$ and $B_k(\mathbf{W}) = \sum_{k'\in \mathcal{K} \backslash k} \lVert \boldsymbol{\theta}_{\rm T}^T \mathbf{H}_{k} \mathbf{w}_{c,k'} \rVert^2 + \lVert \boldsymbol{\theta}_{\rm T}^T \mathbf{H}_{k} \mathbf{W}_{s} \rVert^2 + \sigma_k^2$.
%\begin{align}
%	A_k(\mathbf{W}) &=  \lVert \boldsymbol{\theta}_{\rm T}^T \mathbf{H}_{k} \mathbf{w}_{c,k} \rVert^2, \label{tmp_A} \\
%	B_k(\mathbf{W}) &= \sum_{k'\in \mathcal{K} \backslash k} \lVert \boldsymbol{\theta}_{\rm T}^T \mathbf{H}_{k} \mathbf{w}_{c,k'} \rVert^2 + \lVert \boldsymbol{\theta}_{\rm T}^T \mathbf{H}_{k} \mathbf{W}_{s} \rVert^2 + \sigma_k^2. \label{tmp_B}
%\end{align}
We now transform the fractional part at last term in $\eqref{r_Lan}$ by employing Dinkelbach method \cite{dinkelbach} given by
\begin{align}
	& \frac{ \left( 1+\bar{\gamma}_k^c \right) A_k(\mathbf{W})}{A_k(\mathbf{W}) + B_k(\mathbf{W})} \Rightarrow
	\left( 1+\bar{\gamma}_k^c \right) \Big( A_k( \mathbf{W} ) \!-\!  \lambda_kC_k( \mathbf{W})   \Big),
\end{align}
where $C_k( \mathbf{W}) \triangleq A_k( \mathbf{W}) + B_k( \mathbf{W}) = \sum_{k\in \mathcal{K}} \lVert \boldsymbol{\theta}_{\rm T}^T \mathbf{H}_{k} \mathbf{w}_{c,k} \rVert^2 + \lVert \boldsymbol{\theta}_{\rm T}^T \mathbf{H}_{k} \mathbf{W}_{s} \rVert^2 + \sigma_k^2$, and $\lambda_k = \frac{A_k( \mathbf{W})}{A_k( \mathbf{W}) + B_k( \mathbf{W})}$ is the value obtained at previous iteration. Then the constraint \eqref{con_r} becomes
 \begingroup
 \allowdisplaybreaks
\begin{align}\label{con_r2}
	(1-\eta\xi) & \sum_{k\in\mathcal{K}} \left[ f(\bar{\gamma}_k^c) + \left( 1+\bar{\gamma}_k^c \right) \cdot \Big( A_k( \mathbf{W} ) \!-\! \lambda_k C_k( \mathbf{W}) \Big) \right] \notag\\
	& \qquad\qquad\qquad\qquad - \eta P(\mathbf{W}) \geq 0,
\end{align}
\endgroup
where $f(\bar{\gamma}_k^c) = \log_2(1+ \bar{\gamma}_k^c) - \bar{\gamma}_k^c$.
Similarly, we can have the equivalent constraint for $\eqref{con4}$ as
\begin{align} \label{con4_1}
	f(\bar{\gamma}_k^c) + \left( 1 + \bar{\gamma}_k^c \right) \Big( A_k( \mathbf{W} ) - \lambda_k C_k( \mathbf{W}) \Big)  \geq R_{k,th} .
\end{align}
Furthermore, we reformulate the equivalent constraints in $\eqref{con5}$ and $\eqref{con6}$ respectively as
 \begingroup
 \allowdisplaybreaks
\begin{align} 
	&\lVert \mathbf{u}_{s}^H \mathbf{H}_{s} \mathbf{W}_{s} \rVert^2 - \gamma^s_{th} \left( \sum_{k\in \mathcal{K}} \lVert \mathbf{u}_{s}^H \mathbf{H}_{s} \mathbf{w}_{c,k} \rVert^2  + \sigma_s^2 \lVert \mathbf{u}_{s}\rVert^2 \right) \geq 0, \label{con5_1} \\
 	&\sum_{k\in \mathcal{K}} \lVert \mathbf{u}_{s}^H \mathbf{H}_{s} \mathbf{w}_{c,k} \rVert^2 - \gamma_{I,th}^s \sigma_s^2 \lVert \mathbf{u}_{s}\rVert^2 \leq 0. \label{con6_1}
\end{align}
\endgroup
To this end, the problem $\eqref{problem_Beam}$ becomes 
\begingroup
 \allowdisplaybreaks
% \begin{subequations} \label{problem_Beam2}
\begin{align} \label{problem_Beam2}
    \mathop{\max}\limits_{\substack{   \mathbf{W} \in\{ \mathbf{w}_{c,k}, \mathbf{W}_{s} \}, \\ \mathbf{u}_s, \eta  }} \eta \quad
     \text{ s.t. }
    \eqref{con0_1}, \eqref{con_r2}, \eqref{con4_1}, \eqref{con5_1}, \eqref{con6_1}.
\end{align}
% \end{subequations}
  \endgroup
We can infer that $A_k(\mathbf{W})\geq 0$ and $\eta\xi \lambda_k C_k(\mathbf{W})\geq 0$ are non-convex in $\eqref{con_r2}$ w.r.t. $\mathbf{W}$. Therefore, exploiting SCA \cite{sca} associated with the first-order Taylor expansion, we can derive the approximated affine functions respectively as
 \begingroup
 \allowdisplaybreaks
\begin{align}
	& A_k(\mathbf{W}) \!\approx\! A_k(\breve{\mathbf{W}}) \!+\! \mathfrak{A}\left[ \left( \nabla_{{\mathbf{W}}} A_k( \breve{\mathbf{W}} ) \right)^H \left(\mathbf{W} \!-\! \breve{\mathbf{W}} \right) \right] \notag \\
	& \qquad\qquad\qquad\qquad\qquad\qquad \qquad\qquad\qquad \triangleq \breve{A}_{k}(\mathbf{W}),  \\
	& C_k(\mathbf{W}) = A_k(\mathbf{W}) + B_k(\mathbf{W}) 
	\approx \breve{A}_{k}(\mathbf{W}) + \breve{B}_{k}(\mathbf{W}) \notag \\
	& \! \triangleq \! \breve{A}_{k}(\mathbf{W}) \!+\! B_k(\breve{\mathbf{W}}) \!+\! \mathfrak{A}\left[ \left( \! \nabla_{{\mathbf{W}}} B_k(\! \breve{\mathbf{W}} ) \right)^H \left(\mathbf{W} \!-\! \breve{\mathbf{W}} \!\right) \right],
\end{align}
\endgroup
where operation $\mathfrak{A}$ is $\mathfrak{A}=\mathfrak{R}\{\cdot\}$ if $\mathbf{W}= \mathbf{w}_{c,k}$ and $\mathfrak{A}={\rm Tr}(\cdot)$ if $\mathbf{W}=\mathbf{W}_s$. Notation $\breve{\mathbf{W}} \in \{ \breve{\mathbf{w}}_{c,k}, \breve{\mathbf{W}}_s \}$ indicates the communication/sensing beamformer solutions obtained at previous iteration, where their gradients are $\nabla_{\mathbf{w}_{c,k}} A_k( \breve{\mathbf{w}}_{c,k} ) = 2 \mathbf{H}_k^H \boldsymbol{\theta}_{\rm T} \boldsymbol{\theta}_{\rm T}^T \mathbf{H}_k  \breve{\mathbf{w}}_{c,k}$ and $\nabla_{\mathbf{W}_{s}} B_k( \breve{\mathbf{W}}_{s} ) = 2 \mathbf{H}_k^H \boldsymbol{\theta}_{\rm T} \boldsymbol{\theta}_{\rm T}^T \mathbf{H}_k  \breve{\mathbf{W}}_{s}$.
% \begingroup
% \allowdisplaybreaks
%\begin{align}
%	\nabla_{\mathbf{w}_{c,k}} A_k( \breve{\mathbf{w}}_{c,k} ) &=  \left. \frac{\partial A_k( {\mathbf{w}}_{c,k} )}{ \partial {\mathbf{w}}_{c,k} } \right|_{\mathbf{w}_{c,k} = \breve{\mathbf{w}}_{c,k}} \notag \\
%	& = 2 \mathbf{H}_k^H \boldsymbol{\theta}_{\rm T} \boldsymbol{\theta}_{\rm T}^T \mathbf{H}_k  \breve{\mathbf{w}}_{c,k}, \\
%	%%%%%%%%%%%%%%%%%%%%%%%%%%%%%%%%%%%%%
%	\nabla_{\mathbf{W}_{s}} B_k( \breve{\mathbf{W}}_{s} ) &=  \left. \frac{\partial B_k( {\mathbf{W}}_{s} )}{ \partial {\mathbf{W}}_{s} } \right|_{\mathbf{W}_{s} = \breve{\mathbf{W}}_{s}} \notag \\
%	& = 2 \mathbf{H}_k^H \boldsymbol{\theta}_{\rm T} \boldsymbol{\theta}_{\rm T}^T \mathbf{H}_k  \breve{\mathbf{W}}_{s}.
%\end{align}
%\endgroup
Accordingly, constraint $\eqref{con_r2}$ now becomes
 \begingroup
 \allowdisplaybreaks
\begin{align} \label{con_r3}
	& \sum_{k\in\mathcal{K}} \left[ (1-\eta\xi) f(\bar{\gamma}_k^c) + (1+\bar{\gamma}_k^c) \Big( \breve{A}_{k}(\mathbf{W}) - \eta\xi {A}_{k}(\mathbf{W}) \right. \notag \\
	& \qquad \left. - \lambda_k {C}_{k}(\mathbf{W}) + \eta \xi \breve{C}_{k}(\mathbf{W}) \Big) \right] -\eta P(\mathbf{W}) \geq 0.
\end{align}
\endgroup
Here, we note that $C_k( \mathbf{W})$ depends on the parameters to be solved, i.e.,  $C_k( \mathbf{W})= \breve{A}_{k}(\mathbf{W}) + {B}_{k}(\mathbf{W})$ if $\mathbf{W}=\mathbf{w}_{c,k}$ and $C_k( \mathbf{W})= {A}_{k}(\mathbf{W}) + \breve{B}_{k}(\mathbf{W})$ if $\mathbf{W}=\mathbf{W}_s$. Following similar process, constraint \eqref{con4_1} becomes
\begin{align} \label{con4_2}
f(\bar{\gamma}_k^c) + \left( 1 + \bar{\gamma}_k^c \right) \Big( \breve{A}_k( \mathbf{W} ) - \lambda_k C_k( \mathbf{W}) \Big)  \geq R_{k,th}.
\end{align}
However, we encounter highly coupled terms in constraints $\eqref{con5_1}$ and $\eqref{con6_1}$. We firstly obtain their respective first-order derivatives w.r.t. $\mathbf{W}_s$ and $\mathbf{u}_s$ in non-convex parts as 
	$ f_1(\breve{\mathbf{W}}_s) \triangleq ( \partial \lVert \mathbf{u}_{s}^H \mathbf{H}_{s} \mathbf{W}_{s} \rVert^2 / \partial \mathbf{W}_{s} ) |_{\mathbf{W}_s = \breve{\mathbf{W}}_s } = 2 \mathbf{H}_s^{H} \mathbf{u}_s \mathbf{u}_s^{H} \mathbf{H}_s \breve{\mathbf{W}}_s$,
	$ f_2(\breve{\mathbf{u}}_s) \triangleq (\partial \lVert \mathbf{u}_{s}^H \mathbf{H}_{s} \mathbf{W}_{s} \rVert^2 / \partial \mathbf{u}_{s}  )|_{\mathbf{u}_s = \breve{\mathbf{u}}_s} = 2 \mathbf{W}_{s}^H \mathbf{H}_{s}^H \breve{\mathbf{u}}_s \mathbf{H}_{s} \mathbf{W}_{s}$, and
	$ (\partial \lVert \mathbf{u}_{s}\rVert^2 /  \partial \mathbf{u}_{s}  )|_{\mathbf{u}_s = \breve{\mathbf{u}}_s} = 2 \breve{\mathbf{u}}_s^H$
% \begingroup
% \allowdisplaybreaks
%\begin{align}
%	& \left. \frac{\partial \lVert \mathbf{u}_{s}^H \mathbf{H}_{s} \mathbf{W}_{s} \rVert^2}{  \partial \mathbf{W}_{s} } \right|_{\mathbf{W}_s = \breve{\mathbf{W}}_s } = 2 \mathbf{H}_s^{H} \mathbf{u}_s \mathbf{u}_s^{H} \mathbf{H}_s \breve{\mathbf{W}}_s 
%	\triangleq f_1(\breve{\mathbf{W}}_s) , \\
%	%%%%%%%%%%%%%%%%%%%%%%%%%%%%%%%%%%%
%	& \left. \frac{\partial \lVert \mathbf{u}_{s}^H \mathbf{H}_{s} \mathbf{W}_{s} \rVert^2}{  \partial \mathbf{u}_{s} } \right|_{\mathbf{u}_s = \breve{\mathbf{u}}_s} = 2 \mathbf{W}_{s}^H \mathbf{H}_{s}^H \breve{\mathbf{u}}_s \mathbf{H}_{s} \mathbf{W}_{s}
%	\triangleq f_2(\breve{\mathbf{u}}_s) ,\\
%	%%%%%%%%%%%%%%%%%%%%%%%%%%%%%%%%%%%	
%	& \left. \frac{\partial \lVert \mathbf{u}_{s}\rVert^2}{  \partial \mathbf{u}_{s} } \right|_{\mathbf{u}_s = \breve{\mathbf{u}}_s} = 2 \breve{\mathbf{u}}_s^H.
%\end{align}
%\endgroup
Substituting the above derivatives into $\eqref{con5_1}$ yields two respective constraints w.r.t. $\mathbf{W}_s$ and $\mathbf{u}_s$
 \begingroup
 \allowdisplaybreaks
\begin{subequations} \label{con5_2}
\begin{align}
	&\lVert \mathbf{u}_{s}^H \mathbf{H}_{s} \breve{\mathbf{W}}_{s} \rVert^2 + \tr \left( \left( f_1(\breve{\mathbf{W}}_s) \right) ^H (\mathbf{W}_{s} - \breve{\mathbf{W}}_{s}) \right) \notag \\ &\qquad\qquad\qquad\qquad\qquad\qquad \qquad- \tilde{f}(\mathbf{W},\mathbf{u}_s) \geq 0, \label{con5_2_1} \\
	%%%%%%%%%%%%%%%%%%%%%%%%%%%%%%
	&\lVert \breve{\mathbf{u}}_{s}^H \mathbf{H}_{s} {\mathbf{W}}_{s} \rVert^2 \!+\! \left( f_2(\breve{\mathbf{u}}_s) \right) ^H (\mathbf{u}_{s} - \breve{\mathbf{u}}_{s}) \!-\! \tilde{f}(\mathbf{W},\mathbf{u}_s) \geq 0, \label{con5_2_2}
\end{align}
\end{subequations}
\endgroup
where $\tilde{f}(\mathbf{W},\mathbf{u}_s) = \gamma^s_{th} \left( \sum_{k\in \mathcal{K}} \lVert \mathbf{u}_{s}^H \mathbf{H}_{s} \mathbf{w}_{c,k} \rVert^2  + \sigma_s^2 \lVert \mathbf{u}_{s}\rVert^2 \right)$. Also,
we have the alternative expression for \eqref{con6_1} as
\begin{align} \label{con6_2}
	\sum_{k\in \mathcal{K}} \lVert \mathbf{u}_{s}^H \mathbf{H}_{s} \mathbf{w}_{c,k} \rVert^2 - \gamma_{I,th}^s \sigma_s^2 \left( \lVert \breve{\mathbf{u}}_{s} \rVert^2 + 2\breve{\mathbf{u}}_{s}^H ({\mathbf{u}}_{s} - \breve{\mathbf{u}}_{s}) \right) \leq 0.
\end{align}
We now acquire the equivalent problem as
\begingroup
 \allowdisplaybreaks
% \begin{subequations} \label{problem_Beam3}
\begin{align} \label{problem_Beam3}
    \mathop{\max}\limits_{\substack{   \mathbf{W} =\{ \mathbf{w}_{c,k}, \mathbf{W}_{s} \}, \\ \mathbf{u}_s, \eta  }} \  
    \eta  \quad
     \text{ s.t. } 
    \eqref{con0_1}, \eqref{con_r3}, \eqref{con4_2}, \eqref{con5_2}, \eqref{con6_2}.
\end{align}
% \end{subequations}
  \endgroup
To this end, we can observe that problem \eqref{problem_Beam3} is convex. Then block coordinate descent (BCD) method is applied for acquiring the transmit and receiving ISAC beamforming. The respective convex subproblems can be readily solved by any convex optimization methods, whereas the procedure is elaborated in Algorithm \ref{alg1}. 

\begin{algorithm}[!tb]
  \caption{ Transmit and Receiving ISAC Beamforming}
  \small
  \SetAlgoLined
  \DontPrintSemicolon
  \label{alg1}
  \begin{algorithmic}[1]
   \STATE Initialize $\{\mathbf{w}_{c,k}, \mathbf{W}_s, \mathbf{u}_s, \eta \}$ in problem \eqref{problem_Beam3}
	\REPEAT
		\STATE Solve $\{\mathbf{w}_{c,k}, \eta\}$ with fixed $\{ \breve{\mathbf{W}}_s, \breve{\mathbf{u}}_s \}$
		\STATE Solve $\mathbf{W}_s$ with fixed $\{ \mathbf{w}_{c,k}, \breve{\mathbf{u}}_s, \breve{\eta} \}$
		\STATE Solve $\mathbf{u}_s$ with fixed $\{ \mathbf{w}_{c,k}, \mathbf{W}_s, \breve{\eta} \}$
	\UNTIL{Objective value escalates above a given threshold}
  \end{algorithmic}
\end{algorithm}

%To elaborate a little further, without communication signal interference the conventional solution for receive beamforming for echo signals can be obtained as \cite{?}
%\begin{align}
%	\mathbf{u}_s^* = \frac{\mathbf{H}_s {\rm vec} (\mathbf{W}_s)}{ {\rm vec} (\mathbf{W}_s)^H \mathbf{H}_s^H \mathbf{H}_s  {\rm vec} (\mathbf{W}_s)}.
%\end{align}
%where ${\rm vec}(\mathbf{W}_s)$ vectorize the matrix of $\mathbf{W}_s)$. However, such beamforming may not satisfy the ISAC constraints and will enlarge communication signal interference.

\subsection{STARS with Continuous Configuration}

%\begingroup
% \allowdisplaybreaks
% \begin{subequations} \label{problem_Beam3-1}
%\begin{align}
%    \mathop{\max}\limits_{\substack{  \boldsymbol{\beta}_{\rm T}, \boldsymbol{\beta}_{\rm R},\boldsymbol{\phi}_{\rm T}, \boldsymbol{\phi}_{\rm R} }} &\quad  
%    \eta  \\
%     \text{s.t. } &  \quad
%    \eqref{con1}, \eqref{con4}, \eqref{con5}, \eqref{con6}, \eqref{con0}.
%\end{align}
% \end{subequations}
%  \endgroup
%Firstly, we define a matrix form for phase shifts as $\boldsymbol{\Phi}_{\mathcal{Y}} = \diag (\boldsymbol{\phi}_{\mathcal{Y}}) \in \mathbb{C}^{M\times M}, \forall \mathcal{Y} \in \{\rm R, T \}$. Therefore, we can decouple the STARS configuration as
%\begin{align}
%	\boldsymbol{\theta}_{\mathcal{Y}} = \boldsymbol{\Phi}_{\mathcal{Y}} \boldsymbol{\beta}_{\mathcal{Y}} 
%\end{align}
%To be more precise, we have  \textcolor{red}{unsolvable}
%\begin{align}
%	\boldsymbol{\theta}_{\rm R} &= \boldsymbol{\Phi}_{\rm R} \boldsymbol{\beta}_{\rm R}  \\
%	\boldsymbol{\theta}_{\rm T} &= (\boldsymbol{\Phi}_{\rm R} \pm \boldsymbol{E}_{ \frac{\pi}{2}}  )( \boldsymbol{e} - \boldsymbol{\beta}_{\rm R}^2 )^{\frac{1}{2}}  \\
%\end{align}

\subsubsection{Relaxed STARS}
	The constraint in $\eqref{STAR_amp}$ is relaxed, whereas the T\&R amplitudes are determined independently. We define $|{\theta}_{\mathcal{Y},m}|={\beta}_{\mathcal{Y},m}$ and the problem becomes only related to $\eqref{con4}, \eqref{con5}, \eqref{con6}, \eqref{con0}$,  given by
\begingroup
 \allowdisplaybreaks
 \begin{subequations} \label{problem_Beam3-2}
\begin{align}
    & \mathop{\max}\limits_{\substack{  \boldsymbol{\theta}_{\rm T}, \boldsymbol{\theta}_{\rm R}}} \quad  
    \eta  \\
     & \text{ s.t. }   \quad \
    \beta_{\mathcal{Y},m}\in [0,1], \phi_{\mathcal{Y},m}\in [0, 2\pi], \quad \forall \mathcal{Y}\in \{\rm T, R\},\\
    & \quad\qquad |{\theta}_{\mathcal{Y},m}|={\beta}_{\mathcal{Y},m}, \quad \forall m \in \mathcal{M}, \forall \mathcal{Y}\in \{\rm T, R\}, \label{con_absone} \\
    & \quad\qquad \eqref{con4}, \eqref{con5}, \eqref{con6}, \eqref{con0} \label{new_4cons}.
\end{align}
 \end{subequations}
  \endgroup
We define $A_k(\boldsymbol{\theta}_{\rm T}) =  \lVert \boldsymbol{\theta}_{\rm T}^T \mathbf{H}_{k} \mathbf{w}_{c,k} \rVert^2$ and $C_k( \boldsymbol{\theta}_{\rm T}) = \sum_{k\in \mathcal{K}} \lVert \boldsymbol{\theta}_{\rm T}^T \mathbf{H}_{k} \mathbf{w}_{c,k} \rVert^2 + \lVert \boldsymbol{\theta}_{\rm T}^T \mathbf{H}_{k} \mathbf{W}_{s} \rVert^2 + \sigma_k^2$. We can readily obtain the asymptotic inequality in $\eqref{con_r3}$ with the derivatives of $\nabla_{\boldsymbol{\theta}_{\rm T}} A_k( \breve{\boldsymbol{\theta}}_{\rm T} )= 2 \breve{\boldsymbol{\theta}}_{\rm T}^T \mathbf{H}_k \mathbf{w}_{c,k} \mathbf{w}_{c,k}^H \mathbf{H}_k^H$ and $\nabla_{\boldsymbol{\theta}_{\rm T}} C_k( \breve{\boldsymbol{\theta}}_{\rm T} ) = \sum_{k\in \mathcal{K}} \nabla_{\boldsymbol{\theta}_{\rm T}} A_k( \breve{\boldsymbol{\theta}}_{\rm T} ) + 2 \breve{\boldsymbol{\theta}}_{\rm T}^T \mathbf{H}_k \mathbf{W}_{s} \mathbf{W}_{s}^H \mathbf{H}_k^H$, where $\breve{\boldsymbol{\theta}}_{\rm T}$ is the solution obtained at previous iteration. Also, rate requirement can be acquired in a similar form as that in $\eqref{con4_2}$. The sensing constraints of $\eqref{con5_1}$ and $\eqref{con6_1}$ w.r.t. $\boldsymbol{\theta}_{\rm R}$ are reckoned with the similar manner. However, the diagonal matrix $\boldsymbol{\Theta}_{\rm R}$ in $\mathbf{H}_{s}$ does not have a compact closed-formed derivative. Then it is expressed in a vector form as
%\begin{align*}
%\mathbf{u}_{s}^H \mathbf{H}_{s} \mathbf{W}_{s} & = 
%\mathbf{u}_{s}^H (
%\alpha \mathbf{a}_R(\varphi) \mathbf{a}_T(\varphi)^H + \mathbf{G}^H \boldsymbol{\Theta}_{\rm R} \mathbf{G} ) \mathbf{W}_{s} \\
% 	&= \boldsymbol{\theta}_{\rm R}^T \diag (\mathbf{u}_{s}^H \mathbf{G}^H) \mathbf{G} \mathbf{W}_{s} + \mathbf{c}_s 
% 	\triangleq \boldsymbol{\theta}_{\rm R}^T \tilde{\mathbf{G}}+ \mathbf{c}_s,
%\end{align*}
\begin{align*}
	\mathbf{u}_{s}^H \mathbf{H}_{s} \mathbf{W}_{s} & = \boldsymbol{\theta}_{\rm R}^T \alpha_0^2 \diag (\mathbf{u}_{s}^H \mathbf{G}^H) \mathbf{G} \mathbf{W}_{s} + \mathbf{c}_s 
 	\triangleq \boldsymbol{\theta}_{\rm R}^T \tilde{\mathbf{G}}+ \mathbf{c}_s, \\
 	\mathbf{u}_{s}^H \mathbf{H}_{s} \mathbf{w}_{c,k} 
 	& = \boldsymbol{\theta}_{\rm R}^T \alpha_0^2 \diag (\mathbf{u}_{s}^H \mathbf{G}^H) \mathbf{G} \mathbf{w}_{c,k} + \bar{\mathbf{c}}_s 
 	\triangleq \boldsymbol{\theta}_{\rm R}^T \bar{\mathbf{G}}_{k} + \bar{\mathbf{c}}_s,
\end{align*}
where $\mathbf{c}_s= \alpha_0 \mathbf{u}_{s}^H \mathbf{g}_s \mathbf{g}_s^H \mathbf{W}_{s}$ and $\bar{\mathbf{c}}_s= \alpha_0 \mathbf{u}_{s}^H \mathbf{g}_s \mathbf{g}_s^H \mathbf{w}_{c,k}$ are the constant irrelevant to $\boldsymbol{\theta}_{\rm R}$. 
%Moreover, we have 
%\begin{align*}
%	\mathbf{u}_{s}^H \mathbf{H}_{s} \mathbf{w}_{c,k} & = 
%\mathbf{u}_{s}^H (
%\alpha \mathbf{a}_R(\varphi) \mathbf{a}_T(\varphi)^H + \mathbf{G}^H \boldsymbol{\Theta}_{\rm R} \mathbf{G} ) \mathbf{w}_{c,k} \\
% 	&= \boldsymbol{\theta}_{\rm R}^T \diag (\mathbf{u}_{s}^H \mathbf{G}^H) \mathbf{G} \mathbf{w}_{c,k} + \bar{\mathbf{c}}_s 
% 	\triangleq \boldsymbol{\theta}_{\rm R}^T \bar{\mathbf{G}}_{k} + \bar{\mathbf{c}}_s,
%\end{align*}
We can observe that the only non-convex term in $\eqref{con5_1}$ is $\lVert \mathbf{u}_{s}^H \mathbf{H}_{s} \mathbf{W}_{s} \rVert^2$, whereas $\eqref{con6_1}$ remains convex. Therefore, we conduct SCA to obtain the similar form to $\eqref{con5_2_1}$ with parameters of $\tilde{f}(\boldsymbol{\theta}_{\rm R})= \tilde{f}(\mathbf{W},\mathbf{u}_s)$ and $f_3(\breve{\boldsymbol{\theta}}_{\rm R}) = 2 (\breve{\boldsymbol{\theta}}_{\rm R}^T \tilde{\mathbf{G}} + \mathbf{c}_s) \tilde{\mathbf{G}}^H$, where $\breve{\boldsymbol{\theta}}_{\rm R}$ is the solution at previous iteration. To this end, constraints $\eqref{con4}, \eqref{con5}, \eqref{con6}$, and $\eqref{con0}$ become 
\begingroup
 \allowdisplaybreaks
\begin{align}
	& f(\bar{\gamma}_k^c) + \left( 1 + \bar{\gamma}_k^c \right) \Big( \breve{A}_k( {\boldsymbol{\theta}}_{\rm T} ) - \lambda_k C_k( {\boldsymbol{\theta}}_{\rm T}) \Big)  \geq R_{k,th}, \label{RIS_con_1}\\
%%%%%%%%%%%%%%%%%%%%%%%%%%%%%%%%%%%%%%%%%%%%
	& \left\lVert \breve{\boldsymbol{\theta}}_{\rm R}^T \tilde{\mathbf{G}}+ \mathbf{c}_s \right\rVert^2 + \left( f_3(\breve{\boldsymbol{\theta}}_{\rm R}) \right) ^H ({\boldsymbol{\theta}}_{\rm R} - \breve{\boldsymbol{\theta}}_{\rm R}) - \tilde{f}({\boldsymbol{\theta}}_{\rm R}) \geq 0, \label{RIS_con_2}\\
%%%%%%%%%%%%%%%%%%%%%%%%%%%%%%%%%%%%%%%%%%%%
	& \sum_{k\in \mathcal{K}} \lVert \boldsymbol{\theta}_{\rm R}^T \bar{\mathbf{G}}_k + \bar{\mathbf{c}}_s \rVert^2 - \gamma_{I,th}^s \sigma_s^2 \lVert \mathbf{u}_{s}\rVert^2 \leq 0, \label{RIS_con_3}\\
%%%%%%%%%%%%%%%%%%%%%%%%%%%%%%%%%%%%%%%%%%%%
	& \sum_{k\in\mathcal{K}} \left[ (1-\eta\xi) f(\bar{\gamma}_k^c) + (1+\bar{\gamma}_k^c) \Big( \breve{A}_{k}({\boldsymbol{\theta}}_{\rm T}) - \eta\xi {A}_{k}({\boldsymbol{\theta}}_{\rm T}) \right. \notag \\
	& \qquad \left. - \lambda_k {C}_{k}({\boldsymbol{\theta}}_{\rm T}) + \eta \xi \breve{C}_{k}({\boldsymbol{\theta}}_{\rm T}) \Big) \right] -\eta P \geq 0,  \label{RIS_con_4}
\end{align}
\endgroup
respectively, which are convex. Furthermore, it reveals that the equality constraint in $\eqref{con_absone}$ is non-convex, which can be solved by penalty-based convex-concave programming (PCCP) procedure \cite{my_dstar}. Then the problem with PCCP becomes
\begingroup
 \allowdisplaybreaks
 \begin{subequations} \label{problem_Beam5}
\begin{align}
    & \mathop{\max}\limits_{\substack{  \boldsymbol{\theta}_{\rm T}, \boldsymbol{\theta}_{\rm R}, a_m, b_m}} \quad  
    \eta - \kappa \sum_{m\in \mathcal{M}} (a_m+b_m)  \\
     & \text{s.t. }   \quad\
     |\theta_{\mathcal{Y},m}|^2 \leq \breve{\beta}_{\mathcal{Y},m} \beta_{\mathcal{Y},m} + a_m, \quad \forall m \in \mathcal{M}, \label{con_absone1} \\
    &  \quad\qquad \mathfrak{R}\{ \breve{\theta}_{\mathcal{Y},m} \theta_{\mathcal{Y},m}\} \geq \beta_{\mathcal{Y},m}^2 - b_m, \quad \forall m \in \mathcal{M}, \label{con_absone2} \\
    &  \quad\qquad \eqref{RIS_con_1}, \eqref{RIS_con_2}, \eqref{RIS_con_3}, \eqref{RIS_con_4}.
\end{align}
 \end{subequations}
  \endgroup
Notation $\kappa$ is the penalty term for the auxiliary parameters $a_m$ and $b_m$ in PCCP. The constraints of $|\theta_{\mathcal{Y},m}|^2 \leq \beta_{\mathcal{Y},m}^2$ and of $|\theta_{\mathcal{Y},m}|^2 \geq \beta_{\mathcal{Y},m}^2$ are equivalent union sets for $|\theta_{\mathcal{Y},m}|^2 = \beta_{\mathcal{Y},m}^2$. Utilizing first-order Taylor approximation to the non-convex parts of above two constraints yields $\eqref{con_absone1}$ and $\eqref{con_absone2}$, respectively. To this end, the problem for relaxed STARS is convex and can be solved via arbitrary convex optimization methods.

\subsubsection{Independent STARS}
	The constraint $\eqref{STAR_amp}$ is considered in independent STARS, which is evidently non-convex due to the presence of a quadratic equality constraint. Here, the problem $\eqref{problem_T}$ is rewritten as
\begingroup
 \allowdisplaybreaks
\begin{align} \label{problem_Beam6}
    \mathop{\max}\limits_{\substack{  \boldsymbol{\theta}_{\rm T}, \boldsymbol{\theta}_{\rm R}}} \  
    \eta \quad \text{ s.t. }  \
\eqref{STAR_amp}, \eqref{con4}, \eqref{con5}, \eqref{con6}, \eqref{con0}.
\end{align}
\endgroup
It can be seen that the equality constraint associated with the sum of T\&R amplitudes can be only solved elementwise as it does not have legitimate matrix representation. As a result, we leverage the PDD framework \cite{thz} by defining the auxiliary parameters
\begin{align} \label{con_f}
	\mathbf{f}_1 &= \boldsymbol{\theta}_{\rm R}^T \tilde{\mathbf{G}}, \quad
	\mathbf{f}_{2,k} = \boldsymbol{\theta}_{\rm R}^T \bar{\mathbf{G}}_k , \quad
	\mathbf{f}_{3,k} = \boldsymbol{\theta}_{\rm T}^T \mathbf{H}_k.
\end{align}
The dual variables associated with the respective terms in $\eqref{con_f}$ are defined as $\boldsymbol{\zeta}_1$, $\boldsymbol{\zeta}_{2,k}$ and $\boldsymbol{\zeta}_{3,k}$, with the common penalty constant $\rho$. Then the problem with the augmented Lagrangian expression in the PDD framework is derived as
\begingroup
 \allowdisplaybreaks
 \begin{subequations} \label{problem_Beam7}
\begin{align}
    & \mathop{\max}\limits_{\substack{  \boldsymbol{\theta}_{\rm T}, \boldsymbol{\theta}_{\rm R}}, \mathbf{f}_1, \mathbf{f}_{2,k}, \mathbf{f}_{3,k}} \quad  
    \eta 
    - F(\mathbf{f}, \boldsymbol{\zeta}) \\
%%%%%%%%%%%%%%%%%%%%%%%%%%%%%%%%%%%%%%%%%%%%%    
    & \qquad \text{s.t. }   \
   \eqref{STAR_amp}, \eqref{con4}, \eqref{con5}, \eqref{con6}, \eqref{con0}, \eqref{con_f},
\end{align}
 \end{subequations}
\endgroup
where
\begingroup
 \allowdisplaybreaks
\begin{align}
	F(\mathbf{f}, \boldsymbol{\zeta}) & \!=\! \frac{1}{2\rho} \left( \left\lVert \mathbf{f}_1 \!-\! \boldsymbol{\theta}_{\rm R}^T \tilde{\mathbf{G}} \!+\! \rho \boldsymbol{\zeta}_1 \right\rVert ^2  
	\!+\! \sum_{k\in \mathcal{K}} \left\lVert \mathbf{f}_{2,k} \!-\! \boldsymbol{\theta}_{\rm R}^T \bar{\mathbf{G}}_{k} \!+\! \rho \boldsymbol{\zeta}_{2,k} \right\rVert ^2 \right. \notag\\
    & \left. \qquad\qquad
    \!+\! \sum_{k\in \mathcal{K}}  \left\lVert \mathbf{f}_{3,k} \!-\! \boldsymbol{\theta}_{\rm T}^T \mathbf{H}_k \!+\! \rho \boldsymbol{\zeta}_{3,k} \right\rVert ^2 \right).
\end{align}
\endgroup
Note that the PDD framework possesses two-stage optimization: Inner loop solves problem $\eqref{problem_Beam7}$ and outer loop updates dual variables and penalty factors. Let us first consider the solution set $\{\mathbf{f}_1, \mathbf{f}_{2,k}, \mathbf{f}_{3,k}\}$ and define $A_k(\mathbf{f}_{3,k}) =  \lVert \mathbf{f}_{3,k} \mathbf{w}_{c,k} \rVert^2$ as well as $C_k(\mathbf{f}_{3,k}) = \sum_{k\in \mathcal{K}} \lVert \mathbf{f}_{3,k} \mathbf{w}_{c,k} \rVert^2 + \lVert \mathbf{f}_{3,k} \mathbf{W}_{s} \rVert^2 + \sigma_k^2$. We can readily derive the similar inequality to $\eqref{con_r3}$ with derivatives of $\nabla_{\mathbf{f}_{3,k}} A_k( \breve{\mathbf{f}}_{3,k} )= 2 \breve{\mathbf{f}}_{3,k} \mathbf{w}_{c,k} \mathbf{w}_{c,k}^H$ and $\nabla_{\mathbf{f}_{3,k}} C_k( \breve{\mathbf{f}}_{3,k} ) = \sum_{k\in \mathcal{K}} \nabla_{\mathbf{f}_{3,k}} A_k( \breve{\mathbf{f}}_{3,k} ) + 2 \breve{\mathbf{f}}_{3,k} \mathbf{H}_k \mathbf{W}_{s} \mathbf{W}_{s}^H \mathbf{H}_k^H$, where $\{ \breve{\mathbf{f}}_{1}, \breve{\mathbf{f}}_{3,k}\}$ are solutions at previous iteration. We now rewrite $\eqref{con5_1}$ and $\eqref{con6_1}$ as $\lVert \mathbf{f}_1 + \mathbf{c}_s \rVert^2 - \gamma^s_{th} \left( \sum_{k\in \mathcal{K}} \lVert \mathbf{f}_{2,k} + \bar{\mathbf{c}}_s \rVert^2  + \sigma_s^2 \lVert \mathbf{u}_{s}\rVert^2 \right) \geq 0$ and $\sum_{k\in \mathcal{K}} \lVert \mathbf{f}_{2,k} + \bar{\mathbf{c}}_s \rVert^2 - \gamma_{I,th}^s \sigma_s^2 \lVert \mathbf{u}_{s}\rVert^2 \leq 0$, respectively. We also define $f_4(\breve{\mathbf{f}}_1) = 2(\breve{\mathbf{f}}_1 + \mathbf{c}_s)$ and $\tilde{f}({\mathbf{f}}_{2,k}) = \gamma^s_{th} \left( \sum_{k\in \mathcal{K}} \lVert \mathbf{f}_{2,k} + \bar{\mathbf{c}}_s \rVert^2  + \sigma_s^2 \lVert \mathbf{u}_{s}\rVert^2 \right)$. By performing the similar approximation process to that in $\eqref{RIS_con_1}$--$\eqref{RIS_con_4}$ based on above definitions, we can derive the convex constraints of $ \eqref{con4}, \eqref{con5}, \eqref{con6}$, and $\eqref{con0}$ respectively as
\begingroup
 \allowdisplaybreaks
\begin{align}
	& f(\bar{\gamma}_k^c) + \left( 1 + \bar{\gamma}_k^c \right) \Big( \breve{A}_k( \mathbf{f}_{3,k} ) - \lambda_k C_k( \mathbf{f}_{3,k}) \Big)  \geq R_{k,th}, \label{STAR1_con_1}\\
%%%%%%%%%%%%%%%%%%%%%%%%%%%%%%%%%%%%%%%%%%%%
	& \left\lVert \breve{\mathbf{f}}_1 + \mathbf{c}_s \right\rVert^2 + \left( f_4(\breve{\mathbf{f}}_1) \right) ^H \left( {\mathbf{f}}_1 - \breve{\mathbf{f}}_1 \right) - \tilde{f}({\mathbf{f}}_{2,k}) \geq 0, \label{STAR1_con_2}\\
%%%%%%%%%%%%%%%%%%%%%%%%%%%%%%%%%%%%%%%%%%%%
	& \sum_{k\in \mathcal{K}} \lVert \mathbf{f}_{2,k} + \bar{\mathbf{c}}_s \rVert^2 - \gamma_{I,th}^s \sigma_s^2 \lVert \mathbf{u}_{s}\rVert^2 \leq 0, \label{STAR1_con_3}\\
%%%%%%%%%%%%%%%%%%%%%%%%%%%%%%%%%%%%%%%%%%%%
	& \sum_{k\in\mathcal{K}} \left[ (1-\eta\xi) f(\bar{\gamma}_k^c) + (1+\bar{\gamma}_k^c) \Big( \breve{A}_{k}(\mathbf{f}_{3,k}) - \eta\xi {A}_{k}(\mathbf{f}_{3,k}) \right. \notag \\
	& \qquad\qquad \left. - \lambda_k {C}_{k}(\mathbf{f}_{3,k}) + \eta \xi \breve{C}_{k}(\mathbf{f}_{3,k}) \Big) \right] -\eta P \geq 0.  \label{STAR1_con_4}
\end{align}
\endgroup
By employing BCD, we firstly solve the subproblem w.r.t. $\{  \mathbf{f}_1, \mathbf{f}_{2,k}, \mathbf{f}_{3,k}\}$ by converting the problem $\eqref{problem_Beam7}$ to
\begingroup
 \allowdisplaybreaks
% \begin{subequations} \label{problem_Beam8}
\begin{align}\label{problem_Beam8}
    & \mathop{\min}\limits_{\substack{  \mathbf{f}_1, \mathbf{f}_{2,k}, \mathbf{f}_{3,k}}} \  
    F(\mathbf{f}, \boldsymbol{\zeta}) \quad \text{ s.t. }   \
   \eqref{STAR1_con_1}, \eqref{STAR1_con_2}, \eqref{STAR1_con_3}, \eqref{STAR1_con_4},
\end{align}
% \end{subequations}
\endgroup
which is convex and can be solved by any convex optimization methods. After obtaining optimal solutions of $\{  \mathbf{f}_1, \mathbf{f}_{2,k}, \mathbf{f}_{3,k}\}$, the remaining subproblem is represented by
\begingroup
 \allowdisplaybreaks
% \begin{subequations} \label{problem_Beam9}
\begin{align} \label{problem_Beam9}
    & \mathop{\min}\limits_{\substack{  \boldsymbol{\theta}_{\rm T}, \boldsymbol{\theta}_{\rm R}}} \  
    F(\mathbf{f}, \boldsymbol{\zeta}) \quad \text{s.t. }   \
  \beta_{{\rm T},m}^2 + \beta_{{\rm R},m}^2 = 1,  \ \forall m \in \mathcal{M}.
\end{align}
% \end{subequations}
\endgroup
We further expand $\eqref{problem_Beam9}$ by employing the elementwise optimization problem formulated as
\begingroup
 \allowdisplaybreaks
 \begin{subequations} \label{problem_Beam10}
\begin{align}
    & \mathop{\min}\limits_{\substack{  {\beta}_{{\rm T},m}, {\beta}_{{\rm R},m}, \\ {\phi}_{{\rm T},m}, {\phi}_{{\rm R},m}}} \quad  
	c_{1,m} \beta_{{\rm R}, m}^2 - 2  \beta_{{\rm R}, m}^2 \mathfrak{R}\{ \dot{c}_{2,m} e^{j \phi_{{\rm R},m}} \} \notag \\ 
   & \quad\quad\quad\quad\quad + c_{3,m} \beta_{{\rm R}, m}^2 - 2  \beta_{{\rm R}, m}^2 \mathfrak{R}\{ \dot{c}_{4,m} e^{j \phi_{{\rm R},m}} \} \notag \\
   & \quad\quad\quad\quad\quad + c_{5,m} \beta_{{\rm T}, m}^2 - 2  \beta_{{\rm T}, m}^2 \mathfrak{R}\{ \dot{c}_{6,m} e^{j \phi_{{\rm T},m}} \}\\
%%%%%%%%%%%%%%%%%%%%%%%%%%%%%%%%%%%%%%%%%%%%%    
    & \qquad \text{s.t. }   \
  \beta_{{\rm T},m}^2 + \beta_{{\rm R},m}^2 = 1,  \ \forall m \in \mathcal{M}.
\end{align}
 \end{subequations}
\endgroup
The detailed transformation from problem $\eqref{problem_Beam9}$ to $\eqref{problem_Beam10}$, associated with $c_{i,m}, \forall i\in\{1,2,...,6\}$ can be found in Appendix \ref{Appendix1}. Therefore, the optimal phases are acquired as \cite{thz}
\begingroup
 \allowdisplaybreaks
 \begin{subequations}\label{sol_ph}
\begin{align} 
	\phi_{{\rm R},m} &= \angle (c_{2,m} + c_{4,m}), \ \forall m \in \mathcal{M},\\
	\phi_{{\rm T},m} &= \angle c_{6,m},  \ \forall m \in \mathcal{M}.
\end{align}
 \end{subequations}
 \endgroup
Then we employ trigonometry theory for \eqref{problem_Beam10} by defining ${\beta}_{{\rm T},m} = \cos(\chi_m)$ and ${\beta}_{{\rm R},m} = \sin(\chi_m)$. Substituting \eqref{sol_ph} yields the unconstrained subproblem w.r.t. $\{{\beta}_{{\rm T},m}, {\beta}_{{\rm R},m} \}$ as
\begingroup
 \allowdisplaybreaks
\begin{align} \label{problem_Beam11}
    & \mathop{\min}\limits_{\substack{ 0 \leq \chi_{m} \leq \pi/2 }} \quad  
	g(\chi_{m}) \triangleq c_{1,m} \cos^2(\chi_{m}) - 2|c_{2,m}| \cos(\chi_{m})   \notag\\
	& \qquad\qquad\qquad + c_{3,m} \cos^2(\chi_{m}) - 2|c_{4,m}| \cos(\chi_{m})  \notag\\
	& \qquad\qquad\qquad + c_{5,m} \sin^2(\chi_{m}) - 2|c_{6,m}| \sin(\chi_{m})
\end{align}
\endgroup
Such unconstrained problem $g(\chi_{m})$ with a single variable constrained by a continuous domain can be solved by the ternary search method \cite{tern}. According to the optimal phase-shift $\chi_m^*$, we can obtain the optimal T\&R amplitudes ${\beta}_{{\rm T},m}^* = \sin(\chi_m^*)$ and ${\beta}_{{\rm R},m}^* = \cos(\chi_m^*)$.

\subsubsection{Coupled T\&R STARS}
	We now proceed to consider the coupled phase-shift constraint in $\eqref{STAR_ph}$. With the PDD framework, the problem now becomes
\begingroup
 \allowdisplaybreaks
 \begin{subequations} \label{problem_couple1}
\begin{align}
    & \mathop{\min}\limits_{\substack{  \boldsymbol{\theta}_{\rm T}, \boldsymbol{\theta}_{\rm R}, \mathbf{f}_1, \mathbf{f}_{2,k}, \mathbf{f}_{3,k},\\ \mathbf{f}_{\rm T}, \mathbf{f}_{\rm R}}} \quad  
    F(\mathbf{f}, \boldsymbol{\zeta}) + \sum_{ \mathcal{Y}\in \{\rm T, R\}} \frac{1}{2\rho} \left\lVert \mathbf{f}_{\mathcal{Y}} - \boldsymbol{\theta}_{\mathcal{Y}} + \rho \boldsymbol{\zeta}_{4, \mathcal{Y}} \right\rVert ^2  \\
%%%%%%%%%%%%%%%%%%%%%%%%%%%%%%%%%%%%%%%%%%%%%    
    & \ \quad \text{s.t. }   \
    \mathbf{f}_{\mathcal{Y}} = \boldsymbol{\theta}_{\mathcal{Y}}, \quad \forall \mathcal{Y}\in \{\rm T, R\}, \label{new_con_t} \\
   & \quad\qquad \tilde{\beta}_{{\rm T},m}^2 + \tilde{\beta}_{{\rm R},m}^2 = 1,  \quad \forall m \in \mathcal{M}, \label{new_con_amp}\\
   & \quad\qquad	\cos ( \tilde{\phi}_{{\rm T},m} - \tilde{\phi}_{{\rm R},m} ) = 0,  \quad \forall m \in \mathcal{M},\label{new_con_ph} \\
   & \quad\qquad  \eqref{con4}, \eqref{con5}, \eqref{con6}, \eqref{con0}, \eqref{con_f}.
\end{align}
 \end{subequations}
\endgroup
Note that $\mathbf{f}_{\mathcal{Y}} = [\alpha_m \tilde{\beta}_{\mathcal{Y},m} e^{j \tilde{\phi}_{\mathcal{Y},m}},...,\alpha_m \tilde{\beta}_{\mathcal{Y},M} e^{j \tilde{\phi}_{\mathcal{Y},M}} ]^T$ in $\eqref{new_con_t}$ is an auxiliary variable for $\boldsymbol{\theta}_{\mathcal{Y}}$. We optimize the problem $\eqref{problem_couple1}$ with BCD associated with an order of $\{\mathbf{f}_1, \mathbf{f}_{2,k}, \mathbf{f}_{3,k}\}$, $\{ \boldsymbol{\theta}_{\rm T}, \boldsymbol{\theta}_{\rm R}\}$, and $\{\mathbf{f}_{\rm T}, \mathbf{f}_{\rm R} \}$. It can be readily obtained by solving problem $\eqref{problem_Beam8}$ for the optimal solution of the first block $\{\mathbf{f}_1^*, \mathbf{f}_{2,k}^*, \mathbf{f}_{3,k}^*\}$. Secondly, an unconstrained convex optimization problem w.r.t. $\boldsymbol{\theta}_{\mathcal{Y}}$ can be given by 
\begingroup
 \allowdisplaybreaks
\begin{align} \label{problem_couple2}
    & \mathop{\min}\limits_{\substack{  \boldsymbol{\theta}_{\rm T}, \boldsymbol{\theta}_{\rm R}}} \quad  
    %h(\boldsymbol{\theta}_{\mathcal{Y}}) \triangleq 
    \left\lVert \tilde{\mathbf{G}}^T \boldsymbol{\theta}_{\rm R} - \bar{\boldsymbol{\zeta}}_1 \right\rVert ^2  
    + \sum_{k\in \mathcal{K}} \left\lVert \bar{\mathbf{G}}_{k}^T \boldsymbol{\theta}_{\rm R} - \bar{\boldsymbol{\zeta}}_{2,k} \right\rVert ^2 \notag \\
    & \
    + \sum_{k\in \mathcal{K}} \left\lVert \mathbf{H}_k^T \boldsymbol{\theta}_{\rm T} - \bar{\boldsymbol{\zeta}}_{3,k} \right\rVert ^2 
    + \sum_{ \mathcal{Y}\in \{\rm T, R\}} \left\lVert \boldsymbol{\theta}_{\mathcal{Y}} -  \bar{\boldsymbol{\zeta}}_{4, \mathcal{Y}} \right\rVert ^2
\end{align}
\endgroup
where $\bar{\boldsymbol{\zeta}}_1 = (\mathbf{f}_1^*)^T + \rho {\boldsymbol{\zeta}}_1^T$, $\bar{\boldsymbol{\zeta}}_{2,k} = (\mathbf{f}_{2,k}^*)^T + \rho {\boldsymbol{\zeta}}_2^T$, $\bar{\boldsymbol{\zeta}}_{3,k} = (\mathbf{f}_{3,k}^*)^T + \rho {\boldsymbol{\zeta}}_3^T$ and $\bar{\boldsymbol{\zeta}}_{4, \mathcal{Y}} = \mathbf{f}_{\mathcal{Y}}^T + \rho {\boldsymbol{\zeta}}_{4, \mathcal{Y}}^T$. The unconstrained subproblem \eqref{problem_couple2} is convex can be solved directly.
%Setting the first-order derivative of the objective $\eqref{problem_couple2}$ as $\frac{\partial h(\boldsymbol{\theta}_{\mathcal{Y}})}{\partial \mathcal{Y}}=0$ to zero yields the optimal solution $\boldsymbol{\theta}_{\mathcal{Y}}$ as
%\begingroup
%\allowdisplaybreaks
%\begin{subequations} \label{update_theta}
%\begin{align}
%	\boldsymbol{\theta}_{\rm T}^* &\!=\! \left( \sum_{k\in \mathcal{K}} \mathbf{H}_k^H \mathbf{H}_k \!+\! \mathbf{I}_N \right)^{-1} \left( \sum_{k\in \mathcal{K}} \mathbf{H}_k^H \bar{\boldsymbol{\zeta}}_{3,k} \!+\! \bar{\boldsymbol{\zeta}}_{4, {\rm T}} \right),\\
%%%%%%%%%%%%%%%%%%%%%%%%%%%%%%%%%%%%%%%%%%%%
%	\boldsymbol{\theta}_{\rm R}^* &= \left( \tilde{\mathbf{G}}^H \tilde{\mathbf{G}} +  \sum_{k\in \mathcal{K}} \bar{\mathbf{G}}_{k}^H  \bar{\mathbf{G}}_{k} + \mathbf{I}_N \right)^{-1} \notag\\
%	& \qquad\qquad \left( \tilde{\mathbf{G}}^H \bar{\boldsymbol{\zeta}}_{1} + \sum_{k\in \mathcal{K}} \bar{\mathbf{G}}_{k}^H \bar{\boldsymbol{\zeta}}_{2,k} + \bar{\boldsymbol{\zeta}}_{4, {\rm R}} \right).
%\end{align}
%\end{subequations}
%\endgroup
To this end, the problem becomes only related to $\{\mathbf{f}_{\rm T}, \mathbf{f}_{\rm R}\}$ as
\begingroup
 \allowdisplaybreaks
\begin{align} \label{problem_couple3}
    & \mathop{\min}\limits_{\substack{  \mathbf{f}_{\rm T}, \mathbf{f}_{\rm R}}} \
    \sum_{ \mathcal{Y}\in \{\rm T, R\}}  \left\lVert \mathbf{f}_{\mathcal{Y}} + {\boldsymbol{\zeta}}_{\mathcal{Y}} \right\rVert ^2   \quad \text{s.t. }   \ \eqref{new_con_amp}, \eqref{new_con_ph},
\end{align}
\endgroup
where  $ {\boldsymbol{\zeta}}_{ \mathcal{Y}} = \rho \boldsymbol{\zeta}_{4, \mathcal{Y}} - \boldsymbol{\theta}_{\mathcal{Y}}^* $. Both constraints $\eqref{new_con_amp}, \eqref{new_con_ph}$ are non-convex and difficult to be jointly solved. Therefore, following the asymptotic form in \cite{co1}, we have an alternative problem as
\begin{align} \label{problem_couple4}
    & \mathop{\min}\limits_{\substack{  \tilde{\boldsymbol{\beta}}_{\mathcal{Y}}, \tilde{\boldsymbol{\psi}}_{\mathcal{Y}}}} \  
    \sum_{ \mathcal{Y}\in \{\rm T, R\}}  \mathfrak{R}\left\lbrace {\boldsymbol{\zeta}}_{\mathcal{Y}} \diag(\boldsymbol{\alpha}) \diag(\tilde{\boldsymbol{\beta}}_{\mathcal{Y}}) \tilde{\boldsymbol{\psi}}_{\mathcal{Y}} \right\rbrace   \text{s.t. }    \eqref{new_con_amp}, \eqref{new_con_ph}, \notag
\end{align}
where $\tilde{\boldsymbol{\beta}}_{\mathcal{Y}} = [\tilde{\beta}_{\mathcal{Y},1},...,\tilde{\beta}_{\mathcal{Y},M}] ^T$ and $\tilde{\boldsymbol{\psi}}_{\mathcal{Y}} = [e^{j\tilde{\phi}_{\mathcal{Y},1}},...,e^{j \tilde{\phi}_{\mathcal{Y},M}} ] ^T = [\tilde{{\psi}}_{\mathcal{Y},1},...,\tilde{{\psi}}_{\mathcal{Y},M}]^T$. The optimal solutions for $\tilde{\boldsymbol{\psi}}_{\mathcal{Y}}$ with fixed $\tilde{\boldsymbol{\beta}}_{\mathcal{Y}}$
are obtained as \cite{co1}
 \begingroup
 \allowdisplaybreaks
\begin{align} \label{solset}
	&\mathcal{A}_m =\notag \\
	& \left\lbrace \left( e^{j(\pi-\angle \upsilon_{1,m})} , e^{j(\frac{3\pi}{2}-\angle \upsilon_{1,m})} \right), \left( e^{j(\pi-\angle \upsilon_{2,m})} , e^{j(\frac{\pi}{2}-\angle \upsilon_{2,m})} \right) \right\rbrace,
\end{align}
\endgroup
where $\upsilon_{1,m} = \dot{\theta}_{{\rm T},m}' + j \dot{\theta}_{{\rm R},m}'$ and $\upsilon_{2,m} = \dot{\theta}_{{\rm T},m}' - j \dot{\theta}_{{\rm R},m}'$. We define $\boldsymbol{\theta}_{\mathcal{Y}}' = \diag(\tilde{\boldsymbol{\beta}}_{\mathcal{Y}}^H) \boldsymbol{\zeta}_{\mathcal{Y}} = [{\theta}_{\mathcal{Y},1}',...,{\theta}_{\mathcal{Y},M}']^T$. The solution set $\eqref{solset}$ can be compared to obtain the optimal solution for the following problem
\begin{align} \label{problem_psi}
    & \mathop{\min}\limits_{\substack{ \tilde{\psi}_{\mathcal{Y},m} \in \mathcal{A}_m}} \ 
    \sum_{ \mathcal{Y}\in \{\rm T, R\}}  \mathfrak{R}\left\lbrace \dot{\theta}_{\mathcal{Y},m}' \tilde{\psi}_{\mathcal{Y},m}
    \right\rbrace
\end{align}
Then we can acquire the optimal amplitude given the phase-shifts as $\tilde{\beta}_{{\rm T},m}^*  = \sin \omega_m$ and $ \tilde{\beta}_{{\rm R},m}^* = \cos \omega_m,$
%\begin{align} \label{update_beta}	
%\end{align}
where
\begin{align*}
	\omega_m=\left\{  
\begin{aligned}  
	& -\pi/2 - \varsigma_m ,& & \text{if }  \varsigma_m \in [ -\pi, -\pi/2),\\  
	& 0,& & \text{if }  \varsigma_m \in [ -\pi/2, \pi/4),\\  
	& -\pi/2  ,& & \text{otherwise},
\end{aligned}  
\right.
\end{align*}
where $\varsigma_m = {\rm sign}(c_m)\cos^{-1} ( c_m / \sqrt{c_m^2+d_m^2} )$, $\boldsymbol{\theta}_{\mathcal{Y}}'' = \diag(\tilde{\boldsymbol{\psi}}_{\mathcal{Y}}^{*H}) \boldsymbol{\zeta}_{\mathcal{Y}} = [{\theta}_{\mathcal{Y},1}'',...,{\theta}_{\mathcal{Y},M}'']^T$, $c_m = |\dot{\theta}_{{\rm T},m}'' | \cos (\angle \dot{\theta}_{{\rm T},m}'')$ and $d_m = |\dot{\theta}_{{\rm R},m}'' | \sin (\angle \dot{\theta}_{{\rm R},m}'')$. Note that ${\rm sign}(c_m)=1$ if $c_m > 0$, whilst ${\rm sign}(c_m)=0$ if $c_m \leq 0$.

The dual variables $\{\boldsymbol{\zeta}_1, \boldsymbol{\zeta}_{2,k}, \boldsymbol{\zeta}_{3,k}, \boldsymbol{\zeta}_{4,\mathcal{Y}}\}$ are updated when $f_v \leq \epsilon_{th}$, given by \cite{thz}
\begingroup
\allowdisplaybreaks
\begin{subequations} \label{updatess}
\begin{align}
    \boldsymbol{\zeta}_1 &\leftarrow \boldsymbol{\zeta}_1 + \frac{1}{\rho} \left( \mathbf{f}_1 - \boldsymbol{\theta}_{\rm R}^T \tilde{\mathbf{G}} \right), \label{update1} \\
    \boldsymbol{\zeta}_{2,k} &\leftarrow \boldsymbol{\zeta}_{2,k} + \frac{1}{\rho} \left( \mathbf{f}_{2,k} - \boldsymbol{\theta}_{\rm R}^T \bar{\mathbf{G}}_{k}  \right),  \label{update2} \\
   \boldsymbol{\zeta}_{3,k} & \leftarrow
   \boldsymbol{\zeta}_{3,k} +  \frac{1}{\rho} \left( \mathbf{f}_{3,k} - \boldsymbol{\theta}_{\rm T}^T \mathbf{H}_k \right),  \label{update3}\\
    \boldsymbol{\zeta}_{4, \mathcal{Y}} &  \leftarrow \boldsymbol{\zeta}_{4, \mathcal{Y}} + \frac{1}{\rho} \left( \mathbf{f}_{\mathcal{Y}} - \boldsymbol{\theta}_{\mathcal{Y}}  \right), \label{update4}
\end{align}
\end{subequations}
\endgroup
and
\begin{align} \label{violation}
    f_v = \max \left\{
        \begin{aligned}
        \left\lVert \mathbf{f}_1 - \boldsymbol{\theta}_{\rm R}^T \tilde{\boldsymbol{G}} \right\rVert_{\infty},
        \max_{k\in \mathcal{K}} \left\lVert \mathbf{f}_{2,k} - \boldsymbol{\theta}_{\rm R}^T \bar{\boldsymbol{G}}_k \right\rVert_{\infty}, \\
        \max_{k\in \mathcal{K}} \left\lVert \mathbf{f}_{3,k} - \boldsymbol{\theta}_{\rm T}^T \bar{\boldsymbol{H}}_k \right\rVert_{\infty}, 
        \max_{\mathcal{Y}\in \{ \rm T, R\}} \left\lVert \mathbf{f}_{\mathcal{Y}} - \boldsymbol{\theta}_{\mathcal{Y}} \right\rVert_{\infty}, 
        \end{aligned}
        \right\},
    \end{align}
where $\epsilon_{th}$ is the threshold. If $f_v > \epsilon_{th}$, the penalty factor is updated as $\rho\leftarrow c_{\rho} \rho$, with $c_{\rho}\leq 1$ defined as the learning rate. Note that $\eqref{update4}$ is updated only for coupled T\&R STARS. The algorithm is elaborated in Algorithm \ref{alg_theta}.

\begin{algorithm}[!tb]
  \caption{ Solutions of Independent and Coupled  T\&R STARS}
  \SetAlgoLined
  \DontPrintSemicolon
  \small
  \label{alg_theta}
  \begin{algorithmic}[1]
   \STATE Initialize optimization variables      
\REPEAT
   	\STATE \textbf{(Independent STARS)}: Solve $\{ \mathbf{f}_1, \mathbf{f}_{2,k}, \mathbf{f}_{3,k}\}$ in problem $\eqref{problem_Beam8}$
   	\STATE \textbf{(Coupled T\&R STARS)}: Solve $\{ \mathbf{f}_1, \mathbf{f}_{2,k}, \mathbf{f}_{3,k}, \mathbf{f}_{\mathcal{Y}}\}$ in problem $\eqref{problem_couple1}$ 
	\REPEAT
		\FOR{$m \in \mathcal{M}$}
			\STATE{\textbf{(Independent STARS)}}
			\STATE Update $\phi_{\mathcal{Y},m}$ based on $\eqref{sol_ph}$
			\STATE Solve $\chi_m$ in problem $\eqref{problem_Beam11}$
			\STATE Update ${\beta}_{{\rm T},m}^* = \sin(\chi_m^*)$ and ${\beta}_{{\rm R},m}^* = \cos(\chi_m^*)$
			\STATE{\textbf{(Coupled T\&R STARS)}}
			\STATE Solve $\tilde{\psi}_{\mathcal{Y},m}$ in problem $\eqref{problem_psi}$ based on \eqref{solset}
			\STATE Update $\tilde{\beta}_{{\rm T},m}^*  = \sin \omega_m$ and $ \tilde{\beta}_{{\rm R},m}^* = \cos \omega_m$
		\ENDFOR
	\UNTIL{Convergence}
	\STATE Update dual variables in $\eqref{update1}$--$\eqref{update4}$ if $f_v\leq \varepsilon_{th}$; Otherwise, update penalty factor $\rho\leftarrow c_{\rho} \rho$
%	\IF{$f_v\leq \varepsilon_{th}$}
%		\STATE Update dual variables in $\eqref{update1}$--$\eqref{update4}$
%	\ELSE
%		\STATE Update penalty factor $\rho\leftarrow c_{\rho} \rho$
%	\ENDIF
\UNTIL{Objective value escalates above a given threshold}	
  \end{algorithmic}
\end{algorithm}

\subsection{STARS Quantization}

Based on the previously obtained solutions $\{\beta_{\mathcal{Y},m}^*, \phi_{\mathcal{Y},m}^* \}$, we proceed to optimize STARS quantization levels. First of all, we define the quantization function w.r.t. amplitude and phase-shifts respectively as
\begin{align} 
	\beta_{\mathcal{Y},m}^{Q} = \left\lfloor \beta_{\mathcal{Y},m}^* L_{\beta}^a \right\rfloor \cdot \frac{1}{L_{\beta}^a},  \quad 
	\phi_{\mathcal{Y},m}^{Q} = \left\lfloor \frac{\phi_{\mathcal{Y},m}^*}{2\pi}  L_{\phi}^a \right\rfloor \cdot \frac{2 \pi}{L_{\phi}^a},
\end{align}
where $a\in\{re, in, co\}$ and $\mathcal{Y}\in \{\rm T, R\}$. To investigate the impact of quantization levels on EE, we systematically explore a range of quantization configurations for the STARS parameters. Note that high quantization levels of STARS can provide asymptotic rate performance to the continuous cases but require impractical infinite power consumption of PIN diodes. We rewrite \eqref{power_star} as $P^{a}_{\rm ST} = \left\lceil x^a \right\rceil \sum_{m \in \mathcal{M}} \alpha_m P_{\rm PIN} + P_{\rm CIR}$ where $x^{re} \triangleq \log_2 L^{re}$, $x^{in} \triangleq \log_2 L^{in}$ and $x^{co} \triangleq \log_2 L^{co}+1$. Additionally, $x^{a}$ should satisfy the power constraint of \eqref{con3}, i.e., $x^{a} \leq x^{a}_{th} \triangleq \left\lfloor \frac{P^{a}_{th}-P_{\rm CIR}}{\sum_{m\in \mathcal{M}} \alpha_m P_{\rm PIN}} \right\rfloor$. Then we can obtain the original quantization levels $L^a$ based on $x^a$, i.e.,
 \begingroup
 \allowdisplaybreaks 
\begin{subequations} \label{quant}
\begin{align}
	L^{re} &= (L^{re}_{\beta})^2 (L^{re}_{\phi})^2 = 2^{x^{re}}, \ \text{ for relaxed STARS},\\
	L^{in} &= L_{\beta}^{in} (L^{re}_{\phi})^2 = 2^{x^{in}}, \ \text{ for independent STARS},\\
	L^{co} &= L^{co}_{\beta} L^{co}_{\phi} = 2^{x^{co} -1}, \ \text{ for coupled STARS}.
\end{align}
\end{subequations}
\endgroup
It can be seen that \eqref{quant} is expressed as a product $L^a \triangleq L^a_1 \cdot L^a_2$. Accordingly, we enumerate all feasible integer factor pairs of the $x^a$-dependent $L^a$ that satisfies constraint \eqref{con7}. For example, when $L^a=8=2^{x^a}$ given $x^a=3$, the possible integer factor pairs are $(L^a_1,L^a_2) \in \{ (1, 8), (2, 4), (4, 2), (8, 1)\}$. The optimal EE can then be determined based on the following problem
\begin{align} \label{Problem_quan}
	( L^{a*}_{\beta}, L^{a*}_{\phi}) = 
\argmax_{ ( L^a_{\beta}, L^a_{\phi}) \in \mathcal{B}^a} \ R_t/P
\end{align}
where $\mathcal{B}^a = \{  ( L^a_{\beta}, \lfloor L^a /  L^a_{\beta} \rfloor ) | \forall L^a_{\beta}\in {\rm FAC} (L^a), \forall L^{a,min}_{\mathcal{X}} \leq L^{a}_{\mathcal{X}} \leq L^{a,max}_{\mathcal{X}}, \forall x^{a} \leq x^{a}_{th} \}$, where ${\rm FAC} (\cdot)$ represents the operation of integer factorization providing a set of all possible integer factors. Then we can iteratively evaluate all possible combinations until the the optimal set $(L^{a*}_{\beta} , L^{a*}_{\phi})$ is obtained.

\subsection{Solution of On-Off Control}

Here, we determine which STARS element to be switched on or off, associated with the subproblem as
\begin{align} \label{problem_onoff1}
    & \mathop{\max}\limits_{\substack{  \boldsymbol{\alpha}}} \  
    \eta 
    \quad \text{s.t. }  \ \eqref{con3}, \eqref{con4}, \eqref{con5}, \eqref{con6}, \eqref{con8}, \eqref{con0}.
\end{align}
However, this problem involves an binary set of $\alpha_m\in\{0,1\}$ inducing a non-solvable problem. As a result, inspired by \cite{on3}, we propose a \textit{soft} integer relaxation method handling $\eqref{con8}$ as
\begin{align} \label{ineq_a0}
	0 -\epsilon \leq \alpha_{m} \leq 1+\epsilon \ \Leftrightarrow \ 
	\mathbf{0} -\epsilon \preceq  \boldsymbol{\alpha} \preceq \mathbf{1} + \epsilon,
\end{align}
where $\epsilon$ is a small positive constant. Nevertheless, such process may still provoke unsatisfying solutions if directly using the rounding operation recovering to the original binary solution \cite{on3}. As a result, we define an auxiliary constraint as
\begin{align} \label{ineq_a}
	[\boldsymbol{\alpha}]_m (1-[\boldsymbol{\alpha}]_m) \leq \epsilon,
\end{align}
with the corresponding domain approximately bounded by $\alpha_m \leq \frac{1-\sqrt{1-4\epsilon}}{2}$ and $\alpha_m \geq \frac{1+\sqrt{1-4\epsilon}}{2}$. As a result, the original binary solution $\alpha_m\in\{0,1\}$ can be approximated obtained by jointly considering the joint set of $0\leq \alpha_{m} \leq 1$, $\alpha_m\leq0$ and $\alpha_m\geq 1$ if decaying $\epsilon \rightarrow 0$. Similar proof is omitted here and can be found in \cite{on3}. Since $\eqref{ineq_a}$ is non-convex, we perform SCA to $\eqref{ineq_a}$ as
\begin{align} \label{ineq_a2}
	[\boldsymbol{\alpha}]_m - \left[ [\breve{\boldsymbol{\alpha}}]_m^2 + 2[\breve{\boldsymbol{\alpha}}]_m ([\boldsymbol{\alpha}]_m - [\breve{\boldsymbol{\alpha}}]_m ) \right] \leq \epsilon,
\end{align}
where $[\breve{\boldsymbol{\alpha}}]_m$ is obtained at previous iteration. Different from conventional integer relaxation \cite{on3}, we set additional soft boundary $\epsilon$ to smoothly converge to $\{0,1\}$ by updating $\epsilon\leftarrow \epsilon/\sqrt{t}$, where $t$ indicates the iteration. With sufficiently large $t$, we can have $\mathbf{0} \preceq  \boldsymbol{\alpha} \preceq \mathbf{1}$. Then we rewrite the power consumption of STARS in constraint $\eqref{con3}$ as
\begin{align} \label{power_STAR}
	P^{a}_{{\rm ST}}(\boldsymbol{\alpha}) = \boldsymbol{\alpha}^T \boldsymbol{\alpha} N_{\rm PIN}^{a} P_{\rm PIN} + P_{\rm CIR} \leq P^a_{th}.
\end{align}
Moreover, we decompose the coupled parameter $\boldsymbol{\theta}_{\mathcal{Y}}^T= \boldsymbol{\alpha}^T \bar{\boldsymbol{\Theta}}_{\mathcal{Y}}$ 
where $\bar{\boldsymbol{\Theta}}_{\mathcal{Y}} = \diag([\beta_{\mathcal{Y},1} e^{j\phi_{\mathcal{Y},1}},...,\beta_{\mathcal{Y},M} e^{j\phi_{\mathcal{Y},M}}]^T) $ is unrelated to $\boldsymbol{\alpha}$. Similarly, tackling the non-convex constraints with SCA mechanism renders asymptotic forms to $\eqref{RIS_con_1}$--$\eqref{RIS_con_4}$ as
\begingroup
\allowdisplaybreaks
\begin{align}
	& f(\bar{\gamma}_k^c) + \left( 1 + \bar{\gamma}_k^c \right) \Big( \breve{A}_k( {\boldsymbol{\alpha}} ) - \lambda_k C_k( {\boldsymbol{\alpha}}) \Big)  \geq R_{k,th}, \label{STAR_con_1}\\
%%%%%%%%%%%%%%%%%%%%%%%%%%%%%%%%%%%%%%%%%%%%
	& \left\lVert \breve{\boldsymbol{\alpha}}^T \bar{\boldsymbol{\Theta}}_{\rm R} \tilde{\mathbf{G}} \!+\! \mathbf{c}_s \right\rVert^2 \!+\! \left( f_3(\breve{\boldsymbol{\alpha}}) \right) ^H ( \boldsymbol{\alpha}^T  \!-\! \breve{\boldsymbol{\alpha}}){\boldsymbol{\Theta}}_{\rm R} \!-\! \tilde{f}(\boldsymbol{\alpha}) \geq 0, \label{STAR_con_2}\\
%%%%%%%%%%%%%%%%%%%%%%%%%%%%%%%%%%%%%%%%%%%%
	& \sum_{k\in \mathcal{K}} \lVert \boldsymbol{\alpha}^T \bar{\boldsymbol{\Theta}}_{\rm R} \bar{\mathbf{G}}_k + \bar{\mathbf{c}}_s \rVert^2 - \gamma_{I,th}^s \sigma_s^2 \lVert \mathbf{u}_{s}\rVert^2 \leq 0, \label{STAR_con_3}\\
%%%%%%%%%%%%%%%%%%%%%%%%%%%%%%%%%%%%%%%%%%%%
	& \sum_{k\in\mathcal{K}} \left[ (1-\eta\xi) f(\bar{\gamma}_k^c) + (1+\bar{\gamma}_k^c) \Big( \breve{A}_{k}({\boldsymbol{\alpha}}) - \eta\xi {A}_{k}({\boldsymbol{\alpha}}) \right. \notag \\
	& \qquad \left. - \lambda_k {C}_{k}({\boldsymbol{\alpha}}) + \eta \xi \breve{C}_{k}({\boldsymbol{\alpha}}) \Big) \right] -\eta P(\boldsymbol{\alpha}) \geq 0.  \label{STAR_con_4}
\end{align}
\endgroup
Note that $P(\boldsymbol{\alpha}) = \tr (\mathbf{R}_w) + \xi R_t + P_{\rm BS} + P^{a}_{{\rm ST}}(\boldsymbol{\alpha})$. Due to similar definitions, we omit the expressions of ${A}_k( {\boldsymbol{\alpha}} )$, ${C}_k( {\boldsymbol{\alpha}} )$, $f_3( {\boldsymbol{\alpha}} )$ and $\tilde{f}( {\boldsymbol{\alpha}} )$. To this end, we  reformulate the problem as
\begin{align} \label{problem_onoff2}
    & \mathop{\max}\limits_{\substack{  \boldsymbol{\alpha}}} \  
    \eta \quad \text{ s.t. }  \  \eqref{ineq_a0}, \eqref{ineq_a2}, \eqref{power_STAR}, \eqref{STAR_con_1}\text{--}\eqref{STAR_con_4},
\end{align}
which is convex and can be solved by arbitrary convex optimization tools. To summarize, the proposed AQUES scheme in Algorithm \ref{alg_tot} provides the total solution of transmit and receiving ISAC beamforming, STARS amplitudes/phase-shifts, quantization as well as element selection.

\begin{algorithm}[!tb]
  \caption{ The Proposed AQUES Scheme}
  \SetAlgoLined
  \DontPrintSemicolon
  \small
  \label{alg_tot}
  \begin{algorithmic}[1]
   \STATE Initialize optimization parameters
	\REPEAT
		\STATE Active Beamforming: Solve $\{\mathbf{w}_{c,k}, \mathbf{W}_s, \mathbf{u}_s, \eta \}$ in Algorithm \ref{alg1}
		\STATE Solve $\{\boldsymbol{\theta}_{\rm T}, \boldsymbol{\theta}_{\rm R}\}$ in problem \eqref{problem_Beam5} for relaxed STARS \\ or in Algorithm \ref{alg_theta} for independent/coupled T\&R STARS
		\STATE Solve $\{L^a_{\beta}, L^a_{\phi}\}$ in problem $\eqref{Problem_quan}$ for type-$a$ STARS
		\STATE Solve $\{\boldsymbol{\alpha}\}$ in problem $\eqref{problem_onoff2}$
	\UNTIL{Objective value escalates above a given threshold}
  \end{algorithmic}
\end{algorithm}

%\subsection{Convergence Analysis}
%	We analyze the convergence of the proposed AQUES scheme in Algorithm \ref{alg_tot}.

\subsection{Computational Complexity Analysis} \label{cca}

Here, we analyze the computational complexity of the proposed AQUES scheme in Algorithm \ref{alg_tot}. First of all, the complexity order of \textbf{ISAC beamforming} in Algorithm \ref{alg1} is obtained as $\mathcal{O}(T^{w}(K^3 N^3 + N^6 + N^3)) \approx \mathcal{O}(T^{w}(K^3 N^3 + N^6))$, associated with each term for solving blocks of $\{\mathbf{w}_{c,k},\eta\}$, $\mathbf{W}_s$ and $\mathbf{u}_s$. Notation $T^w$ is the iteration upper bound. Secondly, we analyze \textbf{STARS amplitudes/phase-shifts} in different types. The problem \eqref{problem_Beam5} for solving relaxed STARS is in a complexity order of $\mathcal{O}(T^{re}( K^2M + KM^2 + M^3 ))$, where $T^{re}$ is its iteration upper bound. In Algorithm \ref{alg_theta} for solving independent and coupled STARS, the problem \eqref{problem_Beam8} is in a complexity order of $\mathcal{O}(K^3 N^3)$, whereas problem \eqref{problem_couple1} is in a complexity order of $\mathcal{O}(K^3N^3+K^2M^2+M^3)$. The update of amplitude and phase-shifts of independent and coupled STARS are identical and in a complexity order of $\mathcal{O}(M)$. Note that we neglect the order of the desired precision of ternary search for simplicity. Furthermore, the complexity order of updating dual variables in \eqref{updatess} is acquired as $\mathcal{O}(K+N+M)$. By comparing above, the complexity orders of independent and coupled STARS are derived as $\mathcal{O}(T^{in}(K^3 N^3 + M))$ and $\mathcal{O}(T^{co}(K^3N^3 + K^2M^2 + M^3))$, respectively, where $T^{in/co}$ indicate their corresponding iteration upper bounds. Thirdly, as for STARS \textbf{quantization}, we intend to solve \eqref{Problem_quan} by exhaustively finding all $2^{x^{a}}$ possible combinations, where $x^a \leq x^{a}_{th}$. Therefore, the complexity order of STARS quantization can be obtained as $\mathcal{O}(2^{x^{a}_{th}})$. Lastly, as for \textbf{element selection} in problem \eqref{problem_onoff2}, its complexity order is derived as $\mathcal{O}(K^2M+ KM^2+M^3)$. We further note that the runtime is asymptotically proportional to the algorithm complexity order and substantially depends on the computational capability of the processing units.

 \section{Numerical Results}\label{sec_sim}

\begin{table}[!t]
	\centering
	\scriptsize
	\caption {Parameter Setting}
		\begin{tabular}{ll}
			\hline
            System Parameters & Values \\ \hline \hline
Reference pathloss $h_0$  	& $-20$ dB \\
Noise power $\sigma^2_k,\sigma^2_s$    	& $-90,-90$ dBm \\
Maximum BS power $P_{th}$    & $[28, 44]$ dBm \\
Maximum STARS power $P^a_{th}$    & 25 dBm \\
Rate requirement $R_{k,th}$ & 1 bps/Hz \\
Sensing requirement $\gamma^{s}_{th}$,$\gamma^{s}_{I,th}$ & $\{3, 10\}$ dB \\
Distance of BS-target/STARS/user & $\{5, 30, 50\}$ m \\
Target/STARS/user direction & $\{0, 45, 45\}$ deg \\
Rician factor   & 3 \\
Path loss exponent  & 2.2   \\
Decoding processing constant $\xi$  & 0.3 \\
Power consumption per PIN diode $P_{\rm PIN}$ & $0.33 $ mW \\
STARS circuit power consumption $P_{\rm CIR}$   & 0.1 W \\
BS power consumption $P_{\rm BS}$   & 10 W \\    
Quantization threshold range & $[2^1, 2^{15}]$ \\
\hline
		\end{tabular}
	\label{lable_sim} 
\end{table}

%\begin{figure}[!t]
%\centering
%\includegraphics[width=3in]{fig/deploy_02.eps}
%\caption{ The relative distances of the deployed STARS architecture.}
%\label{fig:simu}
%\end{figure}

The performance of AQUES in STARS-ISAC systems is evaluated. The distances between the BS and target/STARS/users are set to $\{5, 30, 50\}$ meters, respectively, with their directions of $\{0, 45, 45\}$ deg relative to the BS. The operating frequency is $f_c=3.5$ GHz. The learning rate constant in PDD is set to $c_{\rho} = 0.5$. Note that $P_{\rm BS}=10$ W depends on the operational power consumption of radio frequency chains, baseband processing, analog-digital converter, low-noise amplifier and other pertinent components at BS \cite{thz,bb}. The power consumption values of STARS circuit and PIN diode are set to $P_{\rm PIN}=0.33$ mW and $P_{\rm CIR} = 0.1$ W \cite{thz}, respectively. Note that for fair comparison amongst three STARS types, we set amplitudes of the relaxed STARS to $\beta_{{\rm T},m}=\beta_{{\rm R},m}=0.5$. The other parameters are listed in Table \ref{lable_sim}.

    \subsection{Convergence Performance}
\begin{figure}[!t]
\centering
\subfigure[]{
    \includegraphics[width=1.6in]{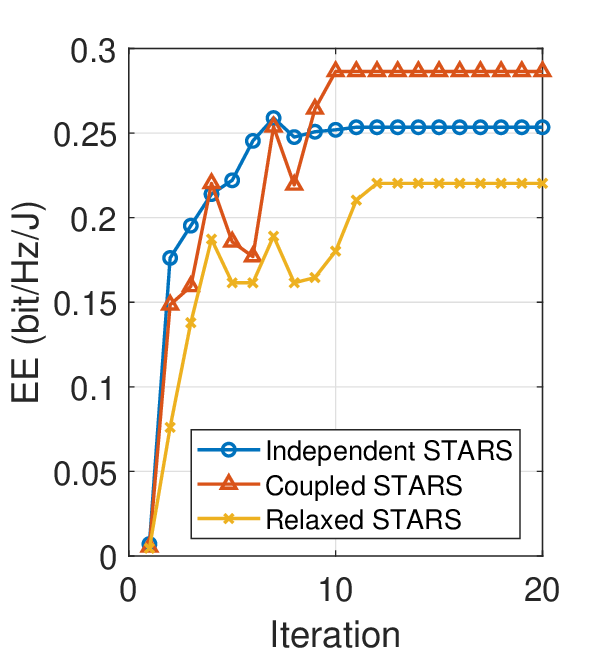}
    \label{fig:f1a}
}
\subfigure[]{
    \includegraphics[width=1.6in]{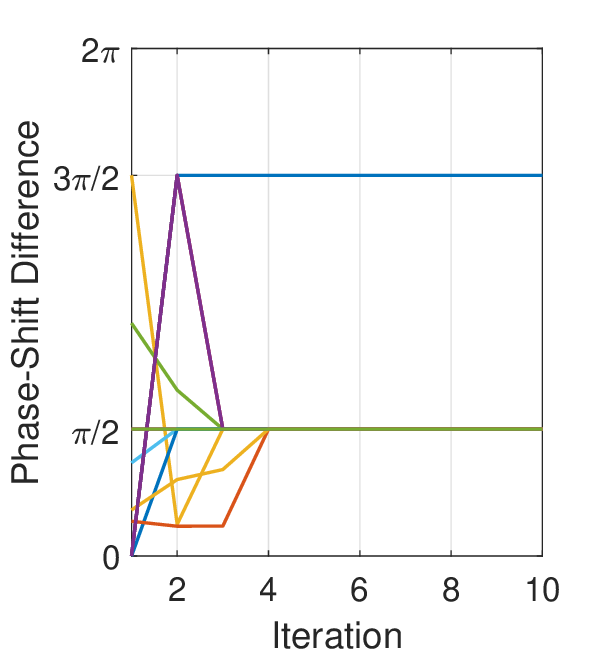}
    \label{fig:f1b}
}
\caption{Convergence of (a) EE for different STARS types and of (b) differences of phase-shifts in coupled STARS.}
\label{fig:f1_ab}
\end{figure}

     Fig.~\ref{fig:f1a} shows convergence of EE performance over iterations for three STARS types: relaxed, independent and coupled STARS. Note that we evaluate all possible quantization combinations in \eqref{Problem_quan} to determine the optimal quantization solution. Consequently, the convergence behavior is primarily influenced by the other three complex constrained subproblems. With alternating optimization, the optimal solution of each subproblem is obtained and then fixed when solving the subsequent subproblems. After each iteration, the updated joint solution achieves improved performance compared to that at previous iteration, as depicted in Fig.~\ref{fig:f1a}. It can be observed that the independent and coupled STARS types converge around the $11$-th iteration, while the relaxed STARS converges slightly later at the $13$-th iteration. Such delayed convergence is attributed to the limited amplitude control range compared to the other STARS types. Moreover, the relaxed STARS requires the largest number of active PIN diodes to configure the elements, which leads to higher power consumption and results in the lowest EE performance. In contrast, the independent STARS governed by \eqref{STAR_amp} utilizes fewer control components than the relaxed STARS and thereby achieves an improved EE. Notably, the coupled STARS attains the highest EE due to its minimal use of STARS components, i.e., the lowest power consumption among the three. Moreover, Fig.~\ref{fig:f1b} exhibits the absolute values of difference of phase-shift $|\phi_{{\rm T},m} - \phi_{{\rm R},m}|$ in coupled T\&R STARS. As observed, the phase-shift differences converge at around the $4$-th iteration to $\frac{\pi}{2}$ or $\frac{3\pi}{2}$, thereby satisfying the constraint $\cos(\phi_{{\rm T},m} - \phi_{{\rm R},m}) = 0$.
    
%%%%%%%%%%%%%%%%%%%%%%%%%%%%%%%%%%%%%%%%%%%%%%%%%%%%%%%%%%%%%%%%%

\begin{figure}[!t]
\centering
\subfigure[]{
    \includegraphics[width=3in]{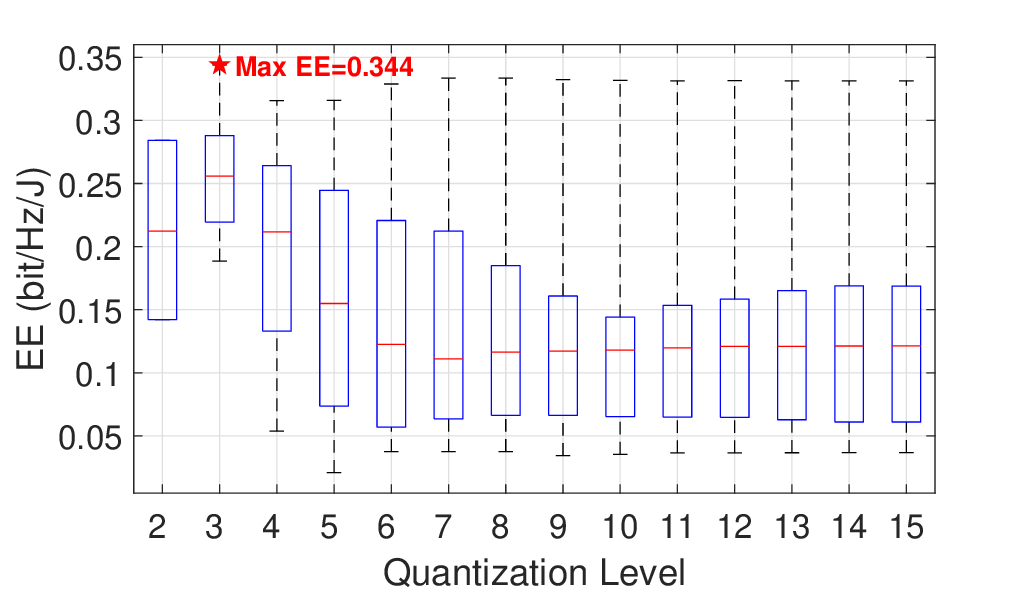}
    \label{fig:boxplot_a}
}
\subfigure[]{
    \includegraphics[width=3in]{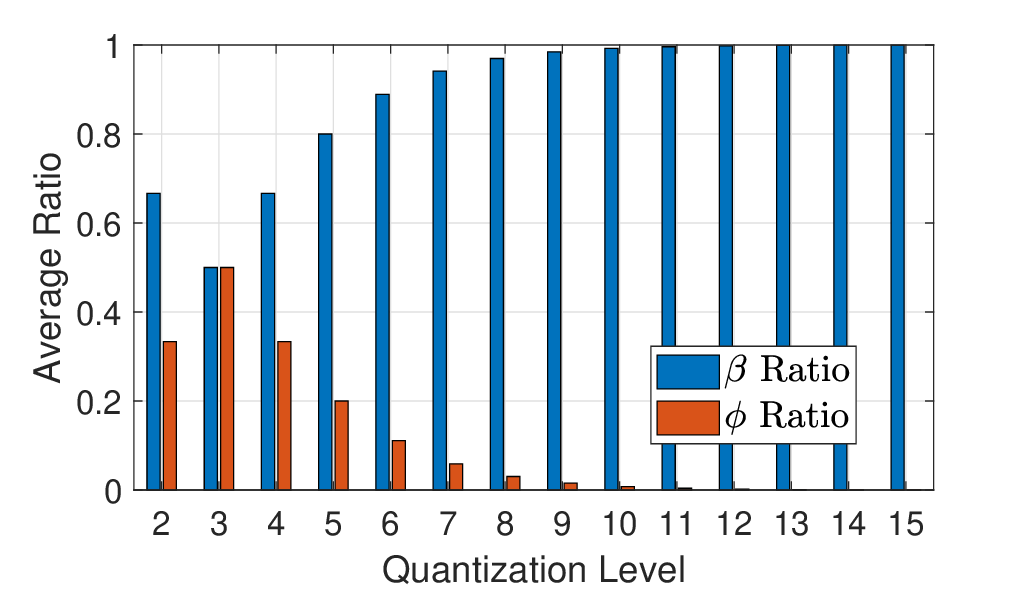}
    \label{fig:boxplot_b}
}
\caption{(a) EE performance under different quantization levels associated with (b) average ratios of amplitudes and phase-shifts.}
\label{fig:boxplot_ab}
\end{figure}

\subsection{Impact of Quantization}

Fig.~\ref{fig:boxplot_ab} presents the EE performance of the coupled STARS under varying quantization levels of $x^{co}\in [2, 15]$ bits. Note that the case with $x^{co}=1$ is excluded as $L^{co} = 2^{1-1}=1$ yields zero amplitudes and zero EE. The case of $x^{co}=15$ provides around $2^{15}=32,768$ discrete regions, which asymptotically approaches the continuous case. The EE performance with all possible combinations of factors are evaluated in a box plot in Fig. \ref{fig:boxplot_a}. It can be observed that the case of $x^{co}=3$ achieves the highest EE of $0.344$ bit/Hz/J, which strikes a compelling tradeoff between the rate and STARS power consumption with a moderate quantization level. For fewer quantization bits of $x^{co}=2$, it has an insufficient configuration resolution, which potentially leads to a lower rate even with the lowest power consumption. However, the cases beyond $x^{co}=3$ will increase not only the rate but also the power consumption of STARS. However, most of solutions under such high-resolution or near-continuous configurations are comparatively lower than those in low-resolution with $x^{co}\leq 4$. Fig.~\ref{fig:boxplot_b} illustrates the average bit allocation ratios between amplitudes $\beta$ and phase-shifts $\phi$ under various quantization levels. The case of $x^{co}=3$ with the highest EE highlights a balanced allocation between $\beta$ and $\phi$, indicating their equal importance at moderate quantization levels. In contrast, other cases exhibit imbalanced allocations, typically assigning more bits to $\beta$ and fewer bits to $\phi$. This trend arises because higher amplitude resolution enables finer power control for rate enhancement, whereas low-resolution phase quantization is often sufficient to steer the beam effectively. For instance, $x^{co} \geq 11$ provides 1-bit phase-shift using $\{0, \pi\}$. Such coarse phase quantization might yield near-optimal EE compared to that of continuous amplitude configurations. To summarize, a balanced and moderate quantization level for amplitude and phase-shift is crucial for high EE performance.

\begin{figure*}[!t]
\centering
\subfigure[]{
    \includegraphics[width=2.2in]{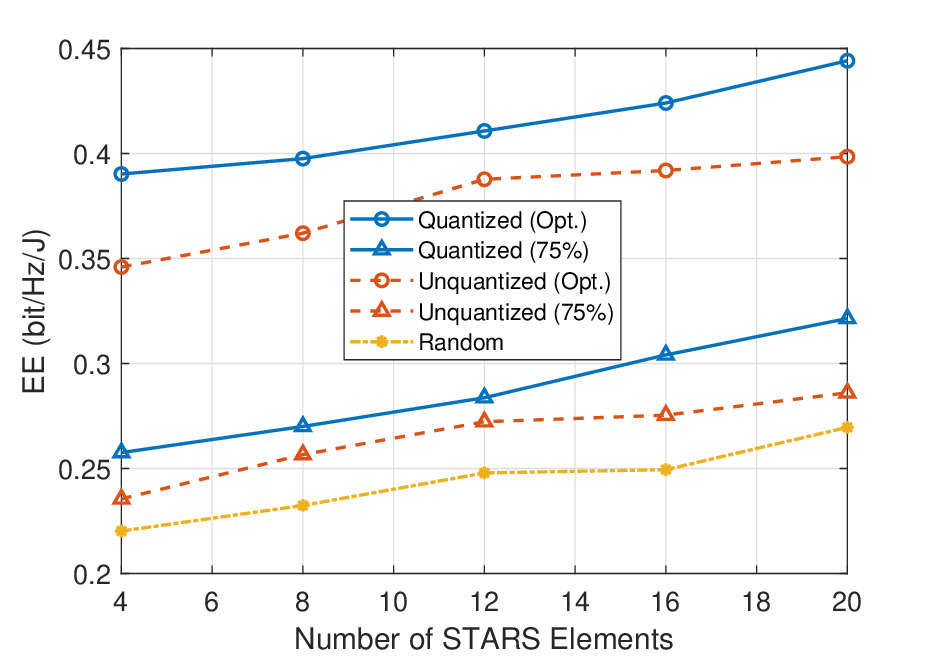}
    \label{fig:f3a}
}
\subfigure[]{
    \includegraphics[width=2.2in]{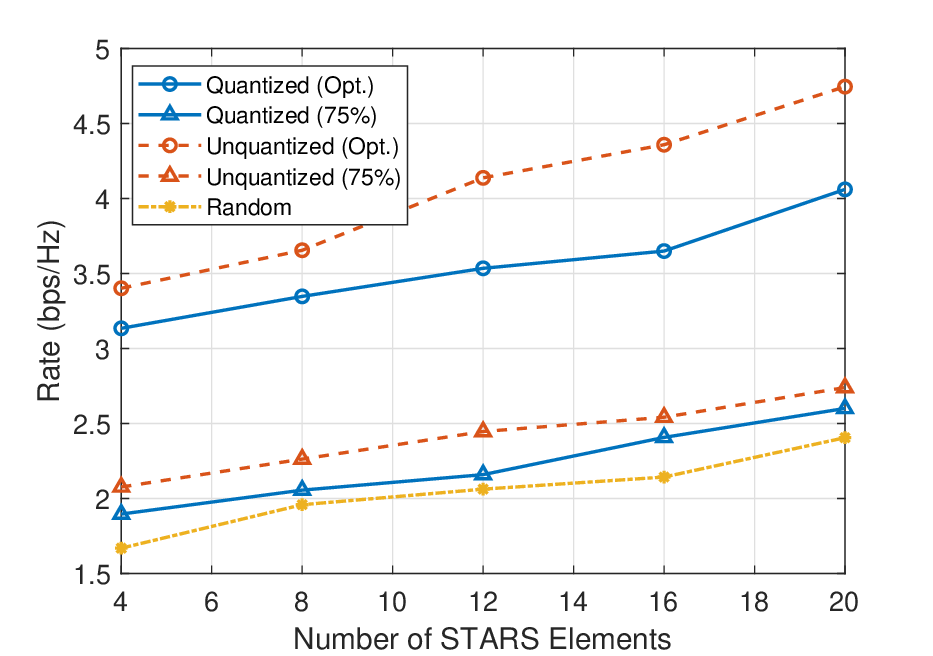}
    \label{fig:f3b}
}
\subfigure[]{
    \includegraphics[width=2.2in]{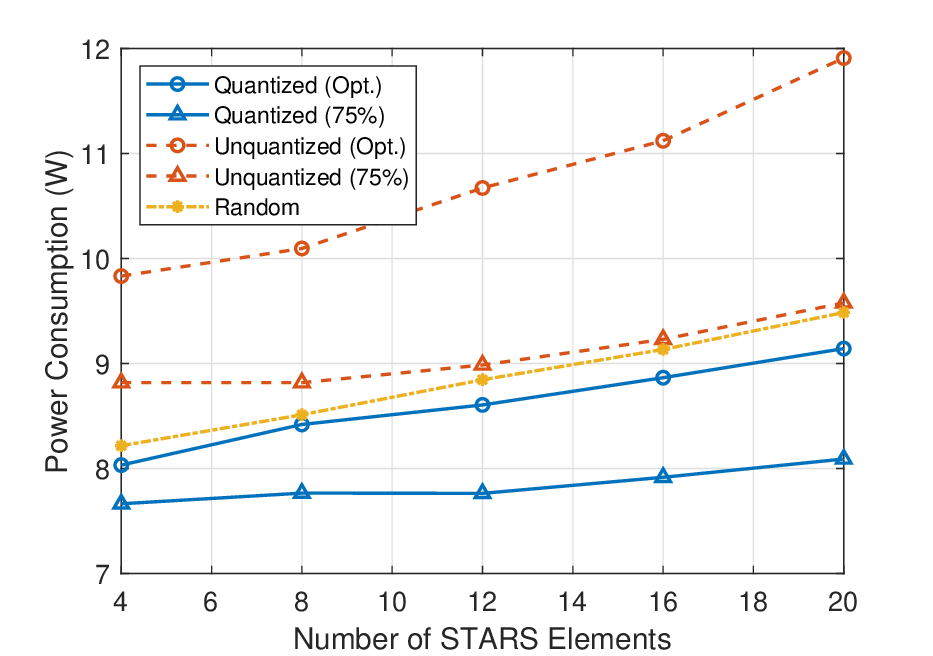}
    \label{fig:f3c}
}
\caption{Performance of (a) EE, (b) rate, and (c) power consumption under different quantization strategies w.r.t. different numbers of STARS elements.}
\label{fig:f3_abc}
\end{figure*}

Fig.~\ref{fig:f3_abc} depicts the performances of EE/rate/power for quantized and unquantized systems under different numbers of STARS elements. Note that optimized quantization is conducted via AQUES scheme, whereas the optimized unquantization indicates near-continuous STARS configurations with $x^{a}=15$. Also, the cases of optimized element selection and fixed $75\%$ of on-elements are compared. Random baseline refers to random quantization with optimized element selection. We can observe that the unquantized STARS has lower EE performance than the quantized one in Fig. \ref{fig:f3a}, as its highest rate is accomplished by consuming the highest power as shown in Figs. \ref{fig:f3b} and \ref{fig:f3c}, respectively. Moreover, random quantization performs the worst EE due to the lowest rate performance. The mechanism under the optimized element selection has an improved EE of around $30\%$ to $40\%$ compared to the fixed case. This is because the optimally selected elements potentially experience better channel quality, providing a higher rate and the corresponding higher EE compared to the fixed case.

%\subsection{Partitioning STARS}
%
%\begin{figure}[!t]
%\centering
%\includegraphics[width=3in]{fig/DSTAR_FIG2.eps}
%\caption{Saperate STARS into two different place.}
%\label{fig:dstar}
%\end{figure}
%
%In fig. \ref{fig:dstar} we partition the original STARS into several sub-
%STARS panels, while having the same total of 40 elements. For example, splitting the STARS into two constituents means that a pair of STARS each having 20 elements is partitioned into two pairs. To evaluate the impact of partitioning on system performance, we further compare three different configurations with varying numbers of antennas $N$. The results show that when STARS is divided into two blocks, EE is effectively improved, showing the benefits of partitioning to the system, which can effectively improve system flexibility. However, when the number of elements in each block exceeds four, the number of elements in each block decreases, and the EE begins to decrease, indicating the loss of beam efficiency due to excessive partitioning.
%The original unpartitioned STARS can achieve the best beamforming gain and energy concentration effect, which is stable but limited by the service range.

\subsection{Different Power Consumption of PIN Diode}

\begin{figure}[!t]
\centering
\includegraphics[width=3in]{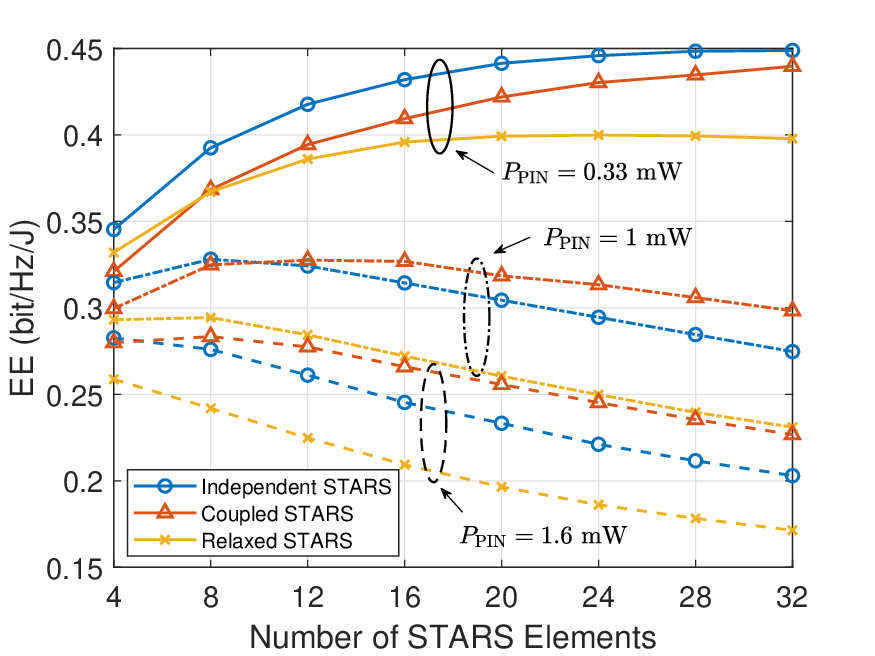}
\caption{EE performance of different STARS types under various numbers of elements and power of PIN diodes.}
\label{fig:IndCouple}
\end{figure}

Fig.~\ref{fig:IndCouple} illustrates the EE performance of independent, coupled and relaxed STARS configurations under varying numbers of STARS elements and different power consumption levels of PIN diode $P_{\rm PIN} = \{0.33, 1, 1.6\}$ mW. We can observe that under low power consumption of $P_{\rm PIN} = 0.33$ mW, the EE improves with increasing numbers of STARS elements amongst all STARS types. In this case, independent STARS outperforms the others due to its flexible control over T\&R phase-shifts, achieving the highest EE of around $0.45$ bit/Hz/J when $M=32$. Coupled STARS also benefits from increasing elements, though with slightly lower EE due to its coupled T\&R phase-shifts. Relaxed STARS exhibits the lowest EE, attributed to its highest power consumption. However, as $P_{\rm PIN}$ increases, EE degrades, particular for the larger numbers of STARS elements. For instance, when $P_{\rm PIN} =\{1, 1.6\}$ mW, STARS elements beyond $M = 12$ leads to declined EE performance. This indicates excessive STARS circuit power consumption from more operating PIN diodes. To elaborate further, it is noteworthy that when $P_{\rm PIN} = \{1, 1.6\}$ mW, the coupled STARS achieves the highest EE performance due to its significantly lower power consumption compared to other configurations. In summary, independent STARS yields the highest EE when hardware cost is low, but its advantage diminishes under higher $P_{\rm PIN}$. The results emphasize the necessity of joint optimization in the number of deployed elements and hardware design in support of high EE performance.

%%%%%%%%%%%%%%%%%%%%%%%%%%%%%%%%%%%%%%%%%%%%

\subsection{Impact of Different ISAC Beamforming}

\begin{figure}[!t]
    \centering
    \includegraphics[width=3in]{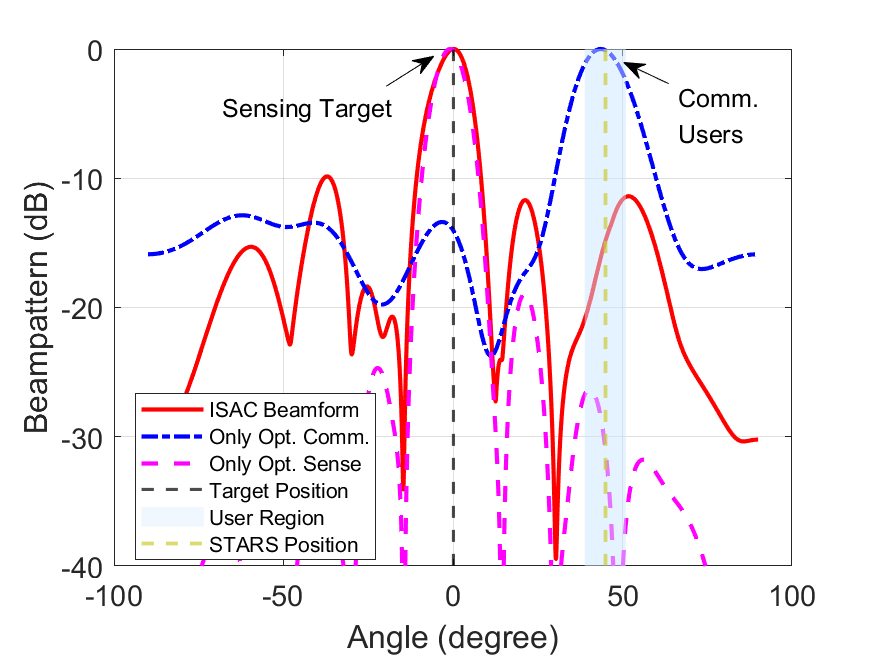}
    \caption{Beampatterns of ISAC and only optimized communication/sensing beamforming under certain target/STARS/user locations.}
    \label{fig:bp}
    \end{figure}

% beampattern
Fig.~\ref{fig:bp} illustrates the beampatterns of the proposed ISAC beamforming with AQUES scheme and the benchmarks of individually optimized communication and sensing beamformers under the given positions of target/STARS/users. It is evident that the ISAC beamforming strikes a balance between communication and sensing objectives. Specifically, the beam exhibits a sharp main lobe aligned with the target position, ensuring high sensing accuracy, while simultaneously maintaining moderate gains within the user region for reliable communications. Additionally, the ISAC beamformer optimized by AQUES achieves the highest normalized beam gain of $0$ dB in the sensing target direction, as additional gain is required to compensate for the power detection loss caused by passive target absorption. In contrast, communication users allocate part of their resources to support sensing, resulting in an ISAC beampattern with a gain of approximately $-12$ dB in communication direction. However, such communication loss can be complemented by the STARS under AQUES. ISAC beamform also adapts to the STARS location as it manipulates the T\&R waveforms to shape the beam effectively toward both sensing and communication directions. In contrast, the communication-only beamformer maximizes its beam gain only within the user region but fails to provide energy at target direction. Similarly, the sensing-only beamformer generates a focused beam at the target direction, yet offers limited signal quality toward the user region.

%%%%%%%%%%%%%%%%%%%%%%%%%%%%%%%%%%%%%%%%%%%%%%%%%%%%%%%%%%%%%%%%%

\begin{figure}[!t]
    \centering
    \includegraphics[width=3in]{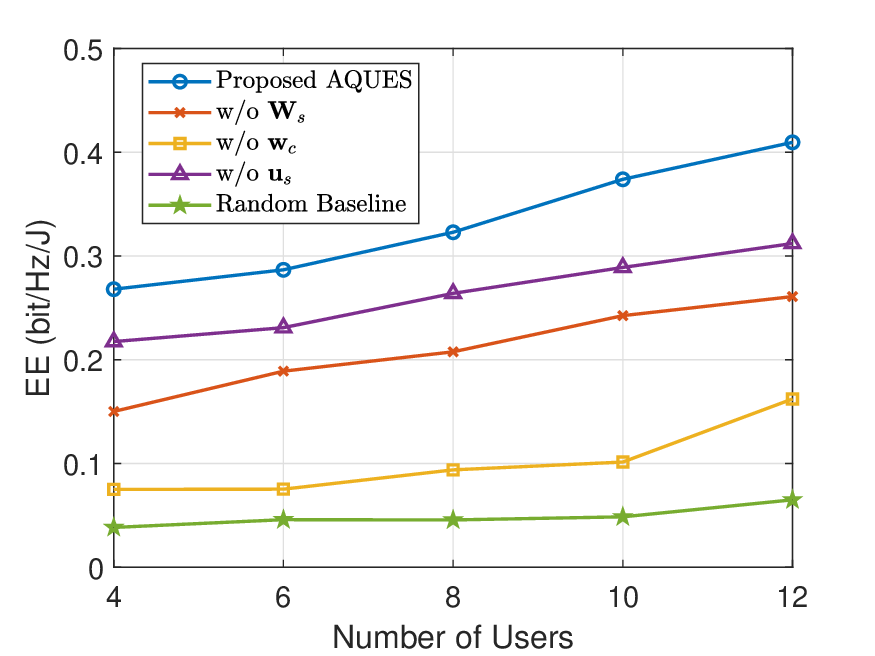}
    \caption{EE of different beamforming schemes with varying numbers of users.}
    \label{fig:bf}
    \end{figure}

    % w/o beamforming 
    Fig.~\ref{fig:bf} demonstrates the EE performance of different beamforming schemes with varying numbers of users, comparing to the cases without optimization for sensing transmit, communication transmit, and sensing receiving beamforming $\{\mathbf{W}_s,\mathbf{w}_c, \mathbf{u}_s\}$ and fully-random beamforming. The proposed AQUES scheme consistently outperforms all baselines, and its EE gain becomes more significant as the number of users increases. This highlights the scalability and effectiveness of AQUES in multi-user STARS-ISAC systems. Among the baselines, removing the sensing transmit beamforming leads to the moderate EE degradation, as it only restricts the flexibility of shaping beams toward the target while maintaining EE for communication users. The case without the receiving sensing beamforming performs slightly better than the case without $\mathbf{W}_s$, indicating that transmit beamforming plays a more dominant role in sensing. However, the absence of communication beamforming $\mathbf{w}_{c}$ leads to a severe degradation in EE. This is due to the critical role of $\mathbf{w}_{c}$ focusing power toward multiple users, thereby influencing both rate and power consumption. Note that the random baseline exhibits the lowest EE due to its lack of coordination in beamforming. To sum up, the results emphasize that the joint design of $\{\mathbf{W}_s,\mathbf{w}_c, \mathbf{u}_s\}$ in AQUES is essential to achieve the highest EE across different numbers of users.

%%%%%%%%%%%%%%%%%%%%%%%%%%%%%%%%%%%%%%%%%%%%%%%%%%%%%%%%%%%%%%%%%

\subsection{STARS Deployment}

\begin{figure}[!t]
\centering
\includegraphics[width=3in]{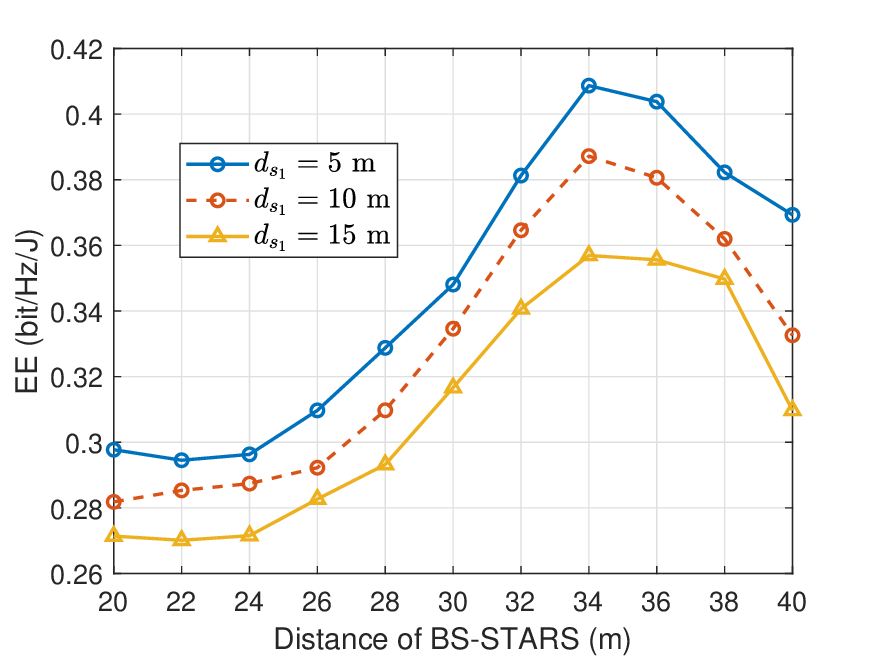}
\caption{EE performance of different distances of BS-STARS and of BS-target.}
\label{fig:distance}
\end{figure}

Fig.~\ref{fig:distance} presents the EE performance versus the distance between the BS and STARS $d$ under different distances between BS and target set to $d_{s_1} =\{5,10,15\}$ m. For all cases, a non-monotonic behavior is observed: EE initially increases with the BS-STARS distance and then gradually declines. The optimal deployment of STARS region with the highest EE is achieved at $d=34$ m. Moreover, a slight degradation of EE occurs when $d\leq 22$ m since some beams may reflect or transmit toward narrow angular ranges, resulting in spatial interference or power inefficiency. When $d_{s_1}$ increases, a lower EE indicates more power is required for sensing the target moving farther away, which is attributed to the increased path loss. This also confirms the importance of spatial alignment among the BS, STARS, and the target. In general, shorter sensing distances allow more flexibility in providing better signal quality to communication users while sustaining the required sensing performance.

%%%%%%%%%%%%%%%%%%%%%%%%%%%%%%%%%%%%%%%%%%%%%%%%%%%%%%%%%%%%%%%%%

\subsection{Analysis of ISAC Power Utilizations}

\begin{figure}[!t]
\centering
\includegraphics[width=3in]{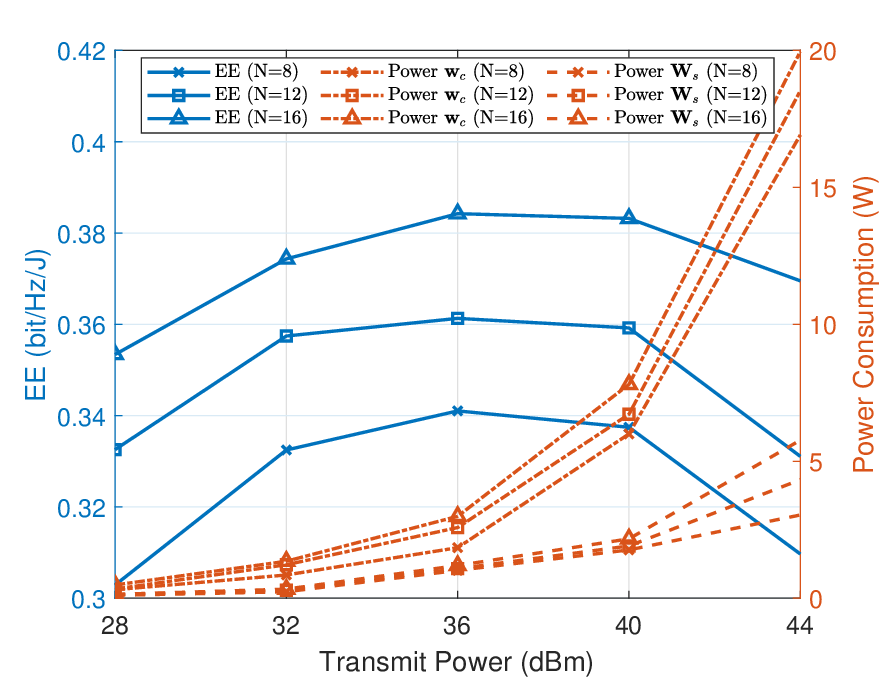}
\caption{Performance of EE and respective power consumption of communication and sensing parts under different transmit power and numbers of antennas.}
\label{fig:pmax_ab}
\end{figure}

Fig.~\ref{fig:pmax_ab} illustrates EE performance and the corresponding power consumption of both communication beamforming $\mathbf{w}_c$ and sensing beamforming $\mathbf{W}_s$ under different transmit power levels and numbers of antennas. We observe that the EE initially increase with transmit power and then decline as power continues to increase. The peaked EE is attained at $P_{th} = 36$ dBm. This trend highlights the fundamental rate-power tradeoff in EE, where increasing transmit power boosts achievable rate, but the marginal rate gains become negligible compared to the dominant power consumption after a certain point. Both communication and sensing power consumption rapidly increases beyond $36$ dBm, leading to the EE decline despite higher rate potential. Furthermore, more antennas provide higher EE, attributed to their improved spatial diversity and beamforming gains. Specifically, $N=16$ antenna configuration achieves the highest EE across all transmit power levels. Intriguingly, the communication power consumption is consistently higher than that of sensing power. This is due to the fact that communication beamforming often requires stronger signal strength across all users, while high-gain of sensing beamforming can tolerate lower used power under certain radar constraints.

%%%%%%%%%%%%%%%%%%%%%%%%%%%%%%%%%%%%%%%%%%%%%%%%%%%%%%%%%%%%%%%%%

\subsection{Benchmark Comparison}

\begin{figure}[!t]
\centering
\subfigure[]{ \includegraphics[width=3in]{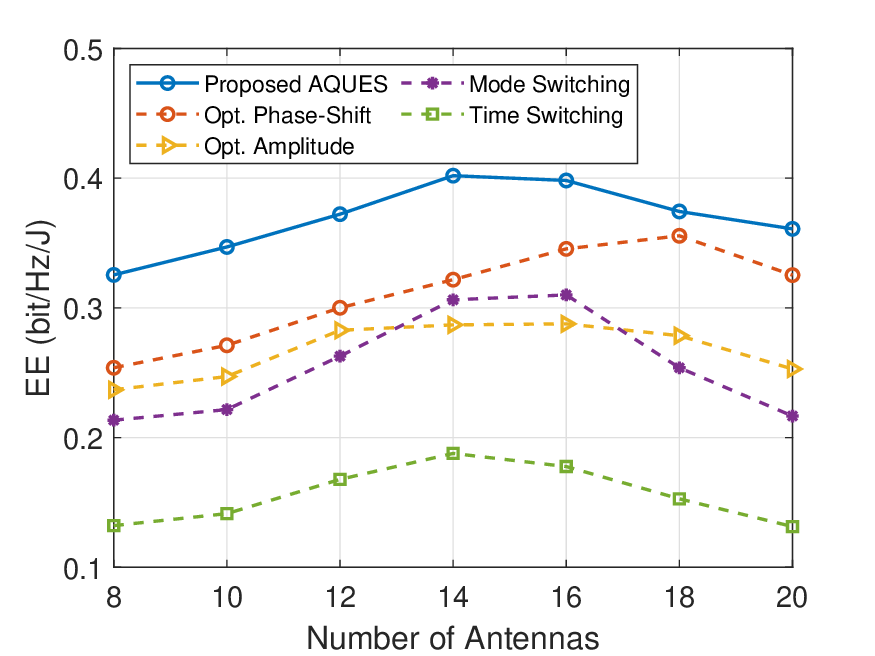} \label{fig:benchmark0}}
\subfigure[]{ \includegraphics[width=3in]{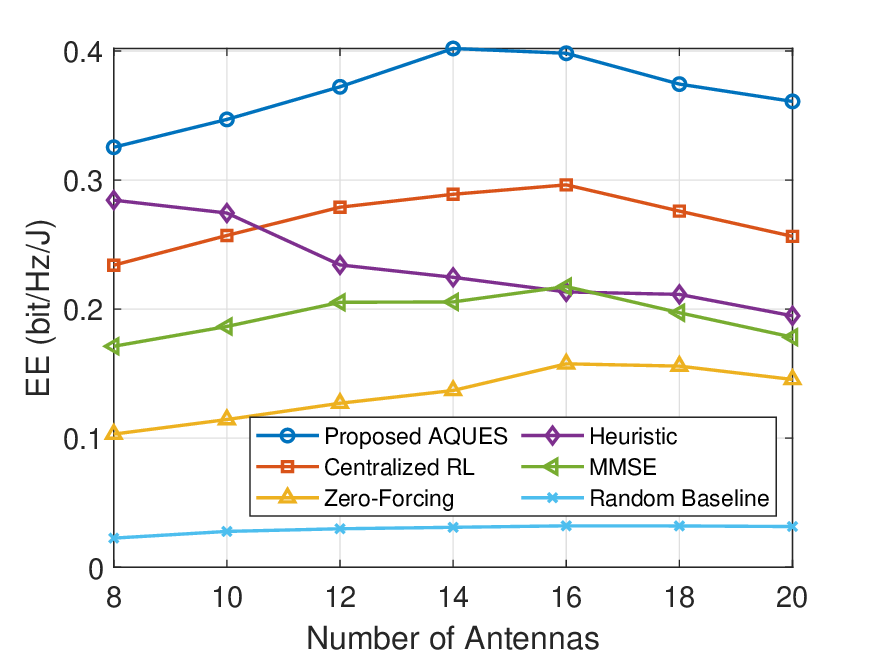}\label{fig:benchmark1} }
\caption{Benchmark comparison of proposed AQUES to (a) different STARS mechanisms and (b) other existing optimization methods.}
\label{fig:BM}
\end{figure}

In Fig. \ref{fig:benchmark0}, we compare the proposed AQUES scheme to the configuration with only optimized either phase-shifts or amplitudes as well as mode switching \cite{ms_case} and time switching \cite{star5} cases. Mode switching indicates that either ${\rm T}$ ($\beta_{{\rm T},m}=1$) or ${\rm R}$ ($\beta_{{\rm R},m}=1$) function is activated per element. Time switching indicates that all elements are operated under either ${\rm T}$ or ${\rm R}$ function. The proposed AQUES scheme under the ES-based STARS scheme outperforms all benchmark methods across different numbers of antennas, achieving the highest EE of approximately 0.4 bit/Hz/J at $N=14$. In contrast, the configuration with only optimized phase-shifts achieves moderate performance, since it fails to leverage available power control by tuning the amplitude. Similarly, the scheme with only optimized amplitudes exhibits limited beamforming flexibility due to the fixed phase-shifts of STARS, resulting in a lower EE than the case with only optimized phase-shifts. The mode switching case associated with a binary selection for T\&R elements yields a lower EE performance most of time compared to the ES-based STARS. This is attributed to its highly constrained beamforming capability and lack of fine-grained control over the STARS. Additionally, the time switching case yields the lowest EE among all scenarios, as only a single function is activated at any given time. Consequently, users can be served and the target can be detected only during partial time slots, which leads to inefficient configurations. Notably, all schemes exhibit a non-monotonic EE trend with respect to the number of antennas. The EE improves as $N$ increases up to around $N=14$ to $16$, but degrades beyond that. This is due to the increased BS power associated with more antennas, which outweighs the marginal rate improvement. To elaborate a little further, mode switching slightly outperforms the optimized amplitude case at $N=14$ and $16$, which can be attributed to the implicit power control and signal separation benefits from the binary selection. Forcing elements to operate in either ${\rm T}$ or ${\rm R}$ mode potentially avoids power splitting losses and can focus energy more decisively toward either the sensing or communication region.

Fig.~\ref{fig:benchmark1} compares AQUES against different benchmarks under varying numbers of BS antennas. Benchmarks are elaborated as follows: 
(1) \textbf{Centralized Reinforcement Learning (RL)} employs Q-learning-based policy \cite{co3} to maximize EE by selecting discrete actions of quantized parameters. The agent interacts with the environment using exploration strategy and updates Q-values based on the observed SINR state and EE reward;
(2) \textbf{Zero-forcing (ZF)} \cite{zero-forcing} is a classical linear precoding technique used to nullify multi-user interference by projecting the transmit signals orthogonally to untargeted users. Normalization is conducted for the final beamformers to satisfy power constraints;
(3) \textbf{Minimum mean square error (MMSE)} \cite{mmse} is designed to jointly balance signal enhancement and interference suppression based on the MMSE criterion. MMSE considers noise statistics and inter-user interference. Normalization is conducted for the final beamformers to satisfy power constraints;
(4) \textbf{Heuristic Genetic Algorithm (GA)} method \cite{ga} is used to search over hybrid discrete-continuous STARS parameters. It leverages population-based elite selection, crossover and mutation operators to explore the optimization space;
(5) \textbf{Random Baseline}: A non-adaptive static scheme with all pertinent parameters randomly generated. 
From Fig.~\ref{fig:benchmark1}, we observe that the proposed AQUES consistently outperforms all benchmarks across all numbers of antennas. Among the baselines, centralized RL performs the second best EE due to the enlarged discrete action space and slow convergence. The heuristic GA-based method shows moderate performance, but it is limited by its local search nature and sensitivity to parameter tuning. MMSE and ZF both exhibit stable but lower EE performances, as they are purely communication-centric beamforming. Note that ZF without interference mitigation capability has a lower EE than MMSE. Finally, the random baseline performs the worst EE, validating the necessity of structured optimization. Regarding the complexity analysis, the proposed AQUES scheme possesses a total complexity order of $\mathcal{O}( K^3 N^3 + N^6 + K^2 M^2 + M^3 + 2^{x^{a}_{th}})$ as analyzed in Subsection \ref{cca}. Centralized RL has a complexity order of $\mathcal{O}(T_{\rm RL} \cdot Q_{\rm c}^{NK} \cdot Q_{\rm s}^N \cdot 2^{2x^{a}_{th}+2M} )$, where $T_{\rm RL}$ is the required iterations and $Q_{\rm c/s}$ is the quantized level of communication/sensing beamforming. The heuristic GA is with an order of $\mathcal{O}(T_{\rm GA} [(N_{\rm GA}+N_{\rm CR}) \cdot (NK + N + 4M) + N_{\rm MU} ])$, where $T_{\rm GA}$ is the required generations and $N_{\rm GA}$ stands for the number of gene solutions. Notations $N_{\rm CR}$ and $N_{\rm MU}$ indicate the number of genes for crossover and the number of gene elements for mutation, respectively. ZF and MMSE methods are with the complexity orders of $\mathcal{O}(KMN+K^2N+K^3)$ and $\mathcal{O}(KMN+N^2K+N^3)$, respectively due to the pseudo-inverse computation of the channel. The random baseline requiring only EE evaluation has a complexity order of $\mathcal{O}(KMN+KN)$. In summary, these results reveal the superiority of the proposed AQUES scheme in adapting to different system parameters configurations, and highlight its robustness compared to existing methods in open literature.

%%%%%%%%%%%%%%%%%%%%%%%%%%%%%%%%%%%%%%%%%%%%%%%%%%%%%%%%%%%%%%%%%
    \section{Conclusion} \label{sec_con}
We have proposed a STARS-ISAC architecture capable of supporting simultaneous downlink multi-user communications and target sensing functionality in a full-space coverage area. We have formulated an EE problem that optimizes the active ISAC beamforming at BS and passive STARS beamforming of amplitudes, phase-shifts, element on-off states and the quantization levels. The resulting non-convex and non-linear problem is solved by utilizing AO algorithm, harnessing Lagrangian dual and Dinkelbach's transformation, PDD framework with PCCP and SCA, as well as heuristic search and integer relaxation mechanisms. Simulation results demonstrate that the proposed STARS-ISAC architecture achieves superior EE performance under coupled T\&R STARS with quantized control. Ablation studies further validate the effectiveness of the ISAC beamforming components and the optimized deployment for STARS. The proposed AQUES scheme outperforms the existing benchmarks, including fixed amplitude/phase-shift, conventional ZF and MMSE beamforming, heuristic GA, centralized learning, and random baselines.

\appendix
\section{Appendix}
\subsection{Transformation from Problem $\eqref{problem_Beam9}$ to $\eqref{problem_Beam10}$} \label{Appendix1}

We reformulate the objective function of problem \eqref{problem_Beam9} into a more tractable form as
\begingroup
\allowdisplaybreaks
\begin{align*}
	& F(\mathbf{f}, \boldsymbol{\zeta}) = \left( \boldsymbol{\theta}_{\rm R}^H \mathbf{\Phi}_1 \boldsymbol{\theta}_{\rm R} - 2 \mathfrak{R} \left\{ \boldsymbol{\theta}_{\rm R}^H \mathbf{v}_1 \right\} \right) \\
	& +  \left( \boldsymbol{\theta}_{\rm R}^H \mathbf{\Phi}_2 \boldsymbol{\theta}_{\rm R} \!-\! 2 \mathfrak{R} \left\{ \boldsymbol{\theta}_{\rm R}^H \mathbf{v}_2 \right\} \right) 
	+ \left( \boldsymbol{\theta}_{\rm T}^H \mathbf{\Phi}_3 \boldsymbol{\theta}_{\rm T} - 2 \mathfrak{R} \left\{ \boldsymbol{\theta}_{\rm T}^H \mathbf{v}_3 \right\} \right) \\
	&
	\triangleq g_1(\boldsymbol{\theta}_{\rm R}) + g_2(\boldsymbol{\theta}_{\rm R}) + g_3(\boldsymbol{\theta}_{\rm T}) + \mathit{C},
\end{align*}
\endgroup
where
$\mathbf{\Phi}_1 = \sum_{k \in \mathcal{K}} \dot{\tilde{\mathbf{G}}}_k \tilde{\mathbf{G}}_k^T$, 
$\mathbf{v}_1 = \sum_{k \in \mathcal{K}} \dot{\tilde{\mathbf{G}}}_k \left( \mathbf{f}_1^T + \rho \boldsymbol{\zeta}_1^T \right)$,
$\mathbf{\Phi}_2 = \sum_{k \in \mathcal{K}} \dot{\bar{\mathbf{G}}}_k \bar{\mathbf{G}}_k^T$,
$\mathbf{v}_2 = \sum_{k \in \mathcal{K}} \dot{\bar{\mathbf{G}}}_k \left( \mathbf{f}_{2,k}^T + \rho \boldsymbol{\zeta}_{2,k}^T \right)$
$\mathbf{\Phi}_3 = \sum_{k \in \mathcal{K}} \dot{\mathbf{H}}_k \mathbf{H}_k^T$, and
$\mathbf{v}_3 = \sum_{k \in \mathcal{K}} \dot{\mathbf{H}}_k \left( \mathbf{f}_{3,k}^T + \rho \boldsymbol{\zeta}_{3,k}^T \right)$. Notation of $\mathit{C}$ represents a constant that does not depend on $\boldsymbol{\theta}_{\mathcal{Y}}$. Note that $\frac{1}{2\rho}$ is neglected here since a constant is unaffected. It is known that $g_{\nu}(\boldsymbol{\theta}_{\mathcal{Y}}), \forall \nu\in\{1,2,3\}$ takes the form of a quadratic function w.r.t. each component of $\boldsymbol{\theta}_{\mathcal{Y}}$. Suppose that $\vartheta_{\mathcal{Y}, m}$ refers to the $m$-th component of $\boldsymbol{\theta}_{\mathcal{Y}}$, we have
\begin{align}\label{g1}
    \tilde{g}_{\nu} (\vartheta_{\mathcal{Y},m}) = c_{\nu_1,m} |\vartheta_{\mathcal{Y},m}|^2 - 2 \mathfrak{R} \left( \dot{c}_{\nu_2,m} \vartheta_{\mathcal{Y},m} \right),
\end{align}
where $ c_{\nu_1,m} , \forall \nu_1 \in \{1,3,5\}$ and $c_{\nu_2,m}, \forall \nu_2 \in \{2,4,6\}$ are complex variables. The exact values of $ c_{\nu_1,m}$ and $c_{\nu_2,m} $ can be obtained via the derivatives as $\frac{\partial \tilde{g}_{\nu}(\vartheta_{\mathcal{Y},m})}{\partial \vartheta_{\mathcal{Y},m}} = c_{\nu_1,m} \vartheta_{\mathcal{Y},m} - \dot{c}_{\nu_2,m}$. Comparing with $\left[ \boldsymbol{\Phi}_{\nu} \boldsymbol{\theta}_{\mathcal{Y}} - \mathbf{v}_{\nu} \right]_m = c_{\nu_1,m} \vartheta_{\mathcal{Y},m} - \dot{c}_{\nu_2,m}$ yields
\begingroup
\allowdisplaybreaks
\begin{subequations}\label{c1toc6}
    \begin{align}
	c_{\nu_1,m} &= [\boldsymbol{\Phi}_{\nu}]_{m,m}, \ \forall (\nu,\nu_1) \in \{(1,1),(2,3),(3,5)\}, \\
	\dot{c}_{2,m} &= [\boldsymbol{\Phi}_1]_{m,m}[\boldsymbol{\theta}_{\rm R}]_m - [\boldsymbol{\Phi}_1 \boldsymbol{\theta}_{\rm R}]_m + [\mathbf{v}_1]_m, \\
	\dot{c}_{4,m} &= [\boldsymbol{\Phi}_2]_{m,m}[\boldsymbol{\theta}_{\rm R}]_m - [\boldsymbol{\Phi}_2 \boldsymbol{\theta}_{\rm R}]_m + [\mathbf{v}_2]_m, \\
	\dot{c}_{6,m} &= [\boldsymbol{\Phi}_3]_{m,m}[\boldsymbol{\theta}_{\rm T}]_m - [\boldsymbol{\Phi}_3 \boldsymbol{\theta}_{\rm T}]_m + [\mathbf{v}_3]_m.
\end{align}
\end{subequations}
\endgroup
Finally, by substituting $\vartheta_{\mathcal{Y},m} = \beta_{\mathcal{Y},m} e^{j \phi_{\mathcal{Y}},m}$ into \eqref{g1}, the auxiliary terms in objective function in $\eqref{problem_Beam10}$ can be obtained. 
%This completes the transformation.

%\section{Appendix}
%\subsection{Ternary Search for Problem \eqref{problem_Beam11}} \label{Appendix2}
%We initially randomly select two points $\{\chi_{m,1},\chi_{m,2}\}$ satisfying $0=l_{min} \leq \chi_{m,1}< \chi_{m,2} \leq l_{max}=\frac{\pi}{2}$. We have three conditions:
%\begin{equation} \label{TS1}
%\left\{  
%\begin{aligned}  
%	&l_{min}\leftarrow \chi_{m,1},& & \text{if }  g(\chi_{m,1}) > g(\chi_{m,2}), \\  
%	&l_{max}\leftarrow \chi_{m,2},& & \text{if } g(\chi_{m,1}) < g(\chi_{m,2}), \\ 
%	&\chi_{m}^* = \frac{\chi_{m,2}+\chi_{m,1}}{2},& & \text{if } g(\chi_{m,1}) = g(\chi_{m,2}). \\ 
%\end{aligned}  
%\right.
%\end{equation}
%The update of $\{\chi_{m,1}, \chi_{m,2}\}$ is
%\begin{align} \label{TS2}
%	\chi_{m,1} & \leftarrow l_{min}+\frac{l_{max}-l_{min}}{3},\\
%	\chi_{m,2} & \leftarrow l_{max}-\frac{l_{max}-l_{min}}{3}. 
%\end{align}
%We terminate the search process when $|\chi_{m,1}-\chi_{m,2}| \leq \chi_{\rm {th}}$ and obtain the optimal $\chi_{m}^* = \frac{\chi_{m,2}+\chi_{m,1}}{2}$ and $g(\chi_{m}^*)$, where $\chi_{\rm {th}}$ indicates the precision threshold.

%\linespread{0.9}
\bibliographystyle{IEEEtran}
\bibliography{IEEEabrv}

\end{document}